\documentclass[
prd,nofootinbib,preprintnumbers,
,twocolumn
]{revtex4}
\usepackage{graphicx,epsf,epsfig}
\usepackage{amsmath, amssymb, amsfonts, bbm}
\usepackage{hyperref}
\usepackage{amssymb}

\usepackage{multirow}
\usepackage{color}

\newcommand{\be}{\begin{equation}}
\newcommand{\ee}{\end{equation}}
\newcommand{\bd}{\begin{displaymath}}
\newcommand{\ed}{\end{displaymath}}
\newcommand{\bea}{\begin{eqnarray}}
\newcommand{\eea}{\end{eqnarray}}
\newcommand{\nn}{\nonumber}

\def\gsim{\raise0.3ex\hbox{$\;>$\kern-0.75em\raise-1.1ex\hbox{$\sim\;$}}}
\def\lsim{\raise0.3ex\hbox{$\;<$\kern-0.75em\raise-1.1ex\hbox{$\sim\;$}}}

\begin{document}

\preprint{IFIC/11-31}  

\title{Soft masses in SUSY SO(10) GUTs with low intermediate scales}

\author{Valentina De Romeri}\email{deromeri@ific.uv.es}
\affiliation{
AHEP Group, Instituto de F\'\i sica Corpuscular -- 
C.S.I.C./Universitat de Val\`encia Edificio de Institutos de Paterna, 
Apartado 22085, E--46071 Val\`encia, Spain}
\author{Martin Hirsch}\email{mahirsch@ific.uv.es}
\affiliation{
AHEP Group, Instituto de F\'\i sica Corpuscular -- 
C.S.I.C./Universitat de Val\`encia Edificio de Institutos de Paterna, 
Apartado 22085, E--46071 Val\`encia, Spain}
\author{Michal Malinsk\'y}\email{malinsky@ific.uv.es}
\affiliation{
AHEP Group, Instituto de F\'\i sica Corpuscular -- 
C.S.I.C./Universitat de Val\`encia Edificio de Institutos de Paterna, 
Apartado 22085, E--46071 Val\`encia, Spain}
\begin{abstract}

The specific shape of the squark, slepton and gaugino mass spectra, if 
measured with sufficient accuracy, can provide invaluable information not 
only about the dynamics underpinning their origin at some very high scale 
such as the unification scale $M_{\rm G}$, but also about the intermediate 
scale physics 
encountered throughout their RGE evolution down to the energy scale 
accessible for the LHC. In this work, we study general features 
of the TeV scale soft SUSY breaking parameters stemming from a generic 
mSugra configuration within certain classes of SUSY $SO(10)$ GUTs with 
different intermediate symmetries below $M_{\rm G}$. We show that 
particular combinations of soft masses show characteristic deviations 
from the mSugra limit in different models and thus, potentially, allow 
to distinguish between these, even if the new intermediate scales are outside 
the energy range probed at accelerators. We also compare our results to 
those obtained for the three minimal seesaw models with mSugra boundary 
conditions and discuss the main differences between those and our 
$SO(10)$ based models.

\end{abstract}

\maketitle
\section{Introduction}

All proposed supersymmetry (SUSY) breaking schemes have to introduce 
some high energy scale, where soft terms are generated. This scale 
could be as high as the scale of Grand unification (GUT), or even 
the Planck scale in gravity 
mediated schemes \cite{Chamseddine:1982jx,Nilles:1983ge}, or as low as 
a $100$ TeV, for example in gauge mediated SUSY breaking (GMSB) 
\cite{Giudice:1998bp}. SUSY particle masses at the electro-weak (TeV) 
scale then have to be calculated from the fundamental parameters of 
the models using renormalization group equations (RGEs).~Although those 
fundamental parameters are a priori unknown, at least in minimal 
schemes there exist certain sum rules for SUSY particle masses, 
which allow to test the different SUSY-breaking mechanisms, as has 
been shown for the example of minimal Supergravity (mSugra) already 
some time ago \cite{Martin:1993ft}. Based on the detailed studies  of 
the capabilities of the LHC and ILC experiments to measure SUSY particle 
masses \cite{AguilarSaavedra:2001rg,Weiglein:2004hn,AguilarSaavedra:2005pw}, 
the accuracy with which different SUSY-breaking schemes can be 
tested has since then been calculated by a number of authors, 
for a few examples see 
\cite{Blair:2000gy,Blair:2002pg,Bechtle:2005vt,Lafaye:2007vs,Adam:2010uz}. 

However, most of these studies concentrated on models in which the
particle spectrum between the electro-weak and the SUSY-breaking scale
was exactly that of the Minimal Supersymmetric extension of the
Standard Model (MSSM).  Evolution under RGEs is, however, sensitive to
the particle content of the theory. Thus, in principle, any superfield
beyond the MSSM (with mass below the SUSY-breaking scale) will leave
its imprint on the soft parameters. The specific shape of the squark,
slepton and gaugino mass spectra, if measured with sufficient
precision, can therefore provide invaluable information not only about
the dynamics underpinning their origin, but also about physics at
intermediate scales.  In this paper, we study soft SUSY-breaking
masses within certain classes of SUSY $SO(10)$ theories with different
intermediate symmetries below the GUT scale $M_{\rm G}$. Our main
motivation to study these models comes from the observed neutrino
masses \cite{Fukuda:1998mi,Ahmad:2002jz,Eguchi:2002dm} and the
possibility that supersymmetry might be discovered soon at the LHC.

In the MSSM, if SUSY particles have TeV-scale masses\footnote{Strictly
speaking, within SUSY unification requires only gauginos to be light
\cite{ArkaniHamed:2004fb,Giudice:2004tc}, but not necessarily
sfermions. We will not entertain this possibility.}, the gauge
couplings unify (nearly) perfectly at around $M_{\rm G} \sim 2\times
10^{16}$ GeV.  Adding new particles beyond the MSSM spectrum can
easily spoil this attractive feature and, thus, the requirement of
gauge coupling unification (GCU) imposes a severe constraint on SUSY
model building. However, neutrino oscillation experiments
\cite{Fukuda:1998mi,Ahmad:2002jz,Eguchi:2002dm,KamLAND2007} have shown
that at least two neutrino masses are non-zero
\cite{Schwetz:2008er,Schwetz:2011qt} and at least one neutrino must
have a mass $m_{\rm Atm} \ge 0.05$ eV. If neutrinos are Majorana
particles, this value indicates that the scale of lepton number
violation (LNV), $\Lambda_{\rm LNV}$, can not be larger (but could
potentially be much smaller, see below) than roughly $\Lambda_{\rm
LNV} \sim 10^{15}$ GeV.  This value is significantly below $M_{\rm
G}$.

In the minimal SUSY $SU(5)$ model neutrinos are massless, just as in
the MSSM and for the same reasons. However, it is fairly
straightforward to extend minimal $SU(5)$ to include a seesaw
mechanism which allows to explain the observed smallness of the
neutrino masses. It is well known that, at the renormalizable level,
there are exactly three ways \cite{Ma:1998dn} to do so: (i) Add (at
least two) gauge singlet superfields, i.e. ``right-handed neutrinos'',
this is now usually called type-I seesaw
\cite{Minkowski:1977sc,seesaw,MohSen}; (ii) add a scalar triplet with
$Y=2$ (type-II seesaw) \cite{Schechter:1980gr,Cheng:1980qt}; or (iii)
add (two or more) {\em fermionic} triplets with $Y=0$ (type-III
seesaw) \cite{Foot:1988aq}. For the latter two cases, the successful
unification of the MSSM can only be maintained if these heavy fields
enter in complete $SU(5)$ multiplets.  Thus, within SUSY models, GCU
requires the type-II seesaw to be realized by adding a pair of Higgs
15-plets, while a type-III seesaw can be generated with (at least two)
copies of 24-plets in the matter sector \cite{Ma:1998dn}.

If we require Yukawas to be perturbative anywhere between the seesaw
scale (i.e., $\Lambda_{\rm LNV}$) and the GUT scale $M_{\rm G}$, all
three types of seesaw require $\Lambda_{\rm LNV}$ to be below
$10^{15}$ GeV. If we ask in addition that all gauge couplings remain
perturbative, lower limits on SUSY type-II seesaw of the order of
$\Lambda_{\rm LNV}\gtrsim 10^{7}$ GeV at 1-loop (or $\Lambda_{\rm
LNV}\gtrsim 10^{9}$ GeV at 2-loop) \cite{Esteves:2010ff} result.  For
type-III seesaw, perturbativity puts lower bounds on the seesaw scale
of the order of $\Lambda_{\rm LNV}\gtrsim 10^{13}$ GeV for three
copies of ${\bf 24}$ of $SU(5)$ and around $\Lambda_{\rm LNV}\gtrsim
10^{9}$ GeV for two\footnote{With only one copy of ${\bf 24}$, the
seesaw scale could be lowered as far as the electro-weak
scale. However, with only one ${\bf 24}$, neutrino data can not be
explained unless non-renormalizable operators are added to the model
\cite{Biggio:2010me}.}  copies of ${\bf 24}$. Since type-I seesaw adds
only Standard Model (SM) singlets, there is no {\em lower} limit on
its scale from perturbativity.

Models based on $SO(10)$ \cite{Fritzsch:1974nn} are different from
$SU(5)$ in that they automatically contain the necessary ingredients
to generate non-zero neutrino masses: (i) The spinorial ${\bf 16}$ of
$SO(10)$ contains a complete SM family plus a gauge singlet, i.e. a
right-handed neutrino. In addition, (ii) $U(1)_{B-L}$ is a subgroup of
$SO(10)$. If the $U(1)_{B-L}$ is broken by $SU(2)_{R}$ triplets with
$B-L=2$, a seesaw mechanism of either type-I and/or type-II results
automatically \cite{MohSen}. Alternatively, breaking $U(1)_{B-L}$ by
$SU(2)_{R}$ doublets can give different realizations of the so-called
inverse \cite{Mohapatra:1986bd} and linear \cite{Akhmedov:1995vm}
seesaw schemes.

The $SO(10)$ gauge symmetry can be broken to the SM gauge group in a
variety of ways \cite{Mohapatra:1986uf}. Since our main motivation is
neutrino masses, all breaking chains of interest to us contain a
left-right symmetry (LR) at some stage. SUSY LR models which use
triplets to break $SU(2)_R$, whether using only $B-L=2$ triplets
\cite{Cvetic:1983su,Kuchimanchi:1993jg} or both $B-L=2$ and $B-L=0$
triplets \cite{Aulakh:1997ba,Aulakh:1997fq}, all require that the
scale of $SU(2)_R$ breaking ($v_R$) is close to the GUT scale,
typically $v_R \ge 10^{15}$ GeV from GCU. However, also in non-minimal
versions of triplet-based models, one can not lower $v_R$ arbitrarily,
since one encounters either problems with proton decay or with
perturbativity \cite{Kopp:2009xt}. Allowing for either (a) sizeable
GUT scale threshold corrections, (b) non-renormalizable operators or
(c) adding some carefully chosen particles the authors of
\cite{Majee:2007uv} find a lower limit on $v_R$ of the order of $v_R
\sim 10^{9}$ GeV.

However, the situation is different in models with doublets. It was
shown in \cite{Malinsky:2005bi} that if one breaks $SU(2)_R\times
U(1)_{B-L}$ by means of an $B-L=0$ triplet to $U(1)_R\times
U(1)_{B-L}$ and, subsequently, the $U(1)_R\times U(1)_{B-L}$ symmetry
gets broken down to $U(1)_Y$ by the $Y$-neutral components of
$SU(2)_{R}$ doublets, it is possible to construct models in which the
scale of $U(1)_R\times U(1)_{B-L}$ breaking, $v_{BL}$, can be as low
as TeV.  In \cite{Majee:2007uv} it was demonstrated that even the full
$SU(2)_R$ can be brought down to the electro-weak scale, if only
doublets are used in the symmetry breaking and if some additional
particles are also light. An especially simple variant for a low $v_R$
scale was discussed in \cite{Dev:2009aw}. Here, GCU is maintained for
a TeV scale $SU(2)_R$ with only two requirements: (a) The numbers of
light left and right doublets have to be different and (b) a (pair of)
light coloured $SU(2)_{L}$ singlets needs to be added to the spectrum.

Obviously, all models with additional gauge groups lead to a
potentially very rich phenomenology at the LHC. Current limits on new
$Z'$ (and $W_R$) gauge bosons are very roughly of the order of
$m_{Z'}\gsim (5-6)/g$ TeV ($m_{W_R} \gsim 1$ TeV)
\cite{Nakamura:2010zzi,Aad:2011xp}, with exact numbers depending on
the couplings, so there is ample room for discovery.  One expects that
for $\sqrt{s}=14$ TeV at the LHC limits for $Z'$ bosons will improve
to at least 3 TeV \cite{Basso:2010pe}. A $W_R$ should be discovered at
the LHC up to masses of the order of $4-4.6$ TeV
\cite{Ferrari:2000sp,Gninenko:2006br}, depending on luminosity. 
However, even if the new gauge bosons predicted in the models
\cite{Majee:2007uv,Malinsky:2005bi,Dev:2009aw} are out of reach for
the LHC, sparticle mass spectra will contain indirect hints for these
new scales due to changes in the RGEs, as discussed above.  This
observations is in fact the main motivation for the calculations
presented in this paper.

Within the mSugra framework, one can define certain combinations of
soft parameters, which are independent of the high scale input
parameters at leading order. We will call such combinations ``RGE
invariants''. In \cite{Buckley:2006nv} it was pointed out that these
invariants show a characteristic deviation from their mSugra
expectations, if either a type-II or a type-III seesaw mediators are
added to the MSSM spectrum. Here, we will study these invariants in
different $SO(10)$ based models. We will construct variants of the
models proposed in \cite{Malinsky:2005bi,Dev:2009aw} and will also
consider a completely new model, in which $v_R$ can be brought down to
the electro-weak scale with the help of an intermediate Pati-Salam
scale \cite{Pati:1974yy}. We will show how the RGE invariants
calculated within these models depart from their mSugra values, how
they differ from model to model and, importantly, also differ from the
expectations for the minimal type-II and type-III seesaws. The
invariants are therefore good indicators to distinguish between
different GUT-based SUSY models.

Two comments might be in order at this point. First, our calculations 
rely on the assumption of strict mSugra boundary conditions. In principle, 
invariants can be calculated also in more complicated SUSY-breaking 
schemes, if the SUSY-breaking scale is larger than the mass scale of 
the new states. However, with a total of only four invariants (per 
generation) only SUSY-breaking schemes with very few additional 
parameters will lead to non-trivial consistency tests. Furthermore, 
while the invariants are certainly useful model discriminators, it 
has been shown that quantitatively important 2-loop corrections exist 
for both, the type-II \cite{Hirsch:2008gh} and the type-III seesaw 
\cite{Esteves:2010ff}. A quantitative determination of the new 
intermediate scales will therefore most likely rely on a detailed 
numerical $\chi^2$-analysis of measured SUSY spectra \cite{Hirsch:2011cw}, 
using invariants only as guidance for which models might be interesting 
for further scrutiny.

The rest of the paper is organized as follows. In the next section, we
shall specify four basic $SO(10)$ models of interest paying particular
attention to their potential compatibility with the SM flavour
structure. In section III, we briefly comment on the evolution of the
soft masses in mSugra models and, for completeness, recall the
definitions of the RGE invariants, following essentially the
discussion in \cite{Hirsch:2008gh}.  The results of a simple 
analysis of their sensitivity to the intermediate scales in the four
scenarios considered here are given in Section IV.  Finally we close
with a short discussion and outlook. Some technical details of the
RGEs in models with more than a single abelian gauge factors are
deferred to an Appendix.

\section{Specific SUSY SO(10) GUT models\label{sect:models}}

Let us begin with a detailed specification of the four basic SUSY
$SO(10)$ GUTs which shall be the studied in Sect.~IV. Though all of
them, by construction, accommodate the low-energy measured values of
the gauge couplings, they will in general yield vastly different MSSM
soft spectra whose shapes would strongly depend on the character of
the intermediate symmetries and the scales of their spontaneous
breakdown.

\subsection{General remarks\label{sect:models:generalities}}

In all cases, we demand that the models should be realistic in several
basic aspects and potentially interesting for our scope, namely:

\begin{itemize}
\item {Requirement 1: SUSY $SO(10)$ unification with a sliding
intermediate scale by which we mean that the position of a certain
intermediate scale can be moved over a large energy range whilst the
full compatibility with the electroweak constraints is
maintained. This is a basic practical stipulation in order to be able
to study the scale-dependence of the soft leading-log RGE invariants
in such GUTs over a large range.}
\item {Requirement 2: Renormalizable $SO(10)\to$ MSSM gauge symmetry
breaking - this is namely to have a good grip on the intermediate
scales and the associated thresholds.}
\item {Requirement 3: Potentially realistic fermionic spectra - we
demand that the effective Yukawa structure is rich enough to be able,
at least in principle, to accommodate the low-energy matter-fermionic
spectra and mixing. The sliding nature of the $SU(2)_{R}\times
U(1)_{B-L}$ scale, however, typically calls for a non-canonical seesaw, 
such as inverse \cite{Mohapatra:1986bd} or linear \cite{Akhmedov:1995vm}
seesaw.}
\item {Requirement 4: MSSM Higgs doublet structure suitable for the
implementation of the standard radiative symmetry breaking and also as
a means to get unrelated Yukawa couplings for quarks and
charged leptons.}
\end{itemize}
As to the Requirement 1 above, we shall be namely interested in SUSY
$SO(10)$ models with a sliding $SU(2)_{R}$ breaking scale which would
be assumed to range from as low as several TeV up to essentially the
GUT scale.  From the gauge unification perspective, there are two
basic strategies to devise such Models. In practice:
\begin{itemize}
\item One can attempt to compensate for the departure of the
$b$-coefficients from their ``canonical'' MSSM values (due to the
presence of $W_{R}^{\pm}$ and the $SU(2)_{R}$-breaking Higgs
multiplets in the desert) by other multiplets brought down to the
$SU(2)_{R}$-breaking scale, which would inflict further shifts to the
$b$-coefficients (namely $g_{3}$) in order to compensate for the
genuine low-scale $SU(2)_{R}$ effects. The main advantage of this
approach is that $SU(2)_{R}\times U(1)_{B-L}$ becomes the only
intermediate scale at play, so the $SO(10)$ gauge symmetry is broken
down to the $SU(3)_{c}\times SU(2)_{L}\times U(1)_{Y}$ of the MSSM
in just two steps. The slight complication here is the fact that the
gauge-coupling unification in such a case is not exact, which brings
an extra theoretical uncertainty into the game. \footnote{This 
feature is already present at the MSSM level.}
\item{Alternatively, rather than compensating for the departure of the
$b$-coefficients from their MSSM values due to the $W_{R}^{\pm}$ (and
the associated Higgs multiplets) in the desert, one can take 
advantage of this and invoke an extra intermediate scale such as for
instance $SU(4)_{C}\times SU(2)_{L}\times SU(2)_{R}$ of Pati and
Salam and let it conspire with the $SU(2)_{R}\times U(1)_{B-L}$ so
that the gauge unification is maintained. Though this can be somewhat
more elaborate in practice, the clear advantage of such a scenario is
that one can always devise an exact gauge coupling unification by a
proper adjustment of the Pati-Salam scale.}
\end{itemize}
In both cases, because we have quite a lot of beyond-MSSM dynamics 
in the desert, we expect significant effects of the relevant intermediate 
scale(s) on the shape of the MSSM squark and slepton spectra.

The first strategy above, especially in combination with the other
requirements, is rather restrictive. Indeed, it imposes strict
conditions on the $b$-coefficients in specific models which should
essentially match those of the MSSM up to a uniform
shift. Nevertheless, a variety of potentially realistic models can
still be devised and, in particular, the behaviour of the RGE
invariants in this class of theories can be strongly
model-specific. We shall demonstrate this on a couple of scenarios of
this kind derived from \cite{Dev:2009aw}, c.f., Model~I and Model~II
in section \ref{SU2Rmodels}.

The sensitivity to the intermediate-scales dynamics should be even
more pronounced in the latter class of scenarios with more than a
single such scale at play. This is namely due to the fact that the
extra fields in the desert associated to a higher intermediate
symmetry (e.g., Pati-Salam) tend to affect the soft spectra stronger
than in the former case with an intermediate
$SU(3)_{c}\times SU(2)_{L}\times SU(2)_{R}\times U(1)_{R}$
only. This feature is going to be clearly visible in the specific
model of this kind, c.f., Model~III in section \ref{SU2Rmodels}.

However, a strong dependence of the invariants on the sliding scale
should not be viewed as a generic feature of the SUSY $SO(10)$
GUTs. Indeed, there are simple scenarios in which the sliding
intermediate scale does leave almost no imprints in the soft
spectrum. We shall demonstrate this on a specific model with a sliding
intermediate $U(1)_{B-L}$ scale (and a fixed $SU(2)_{R}$ scale at
around $10^{14}$ GeV ensuring a proper gauge unification) of the kind
given in \cite{Malinsky:2005bi}, c.f., Model~IV in the section
\ref{SU2Rmodels}. Here, the GUT-scale pattern of the RGE invariants is
(almost) not changed by the running, leaving no good handle on the
intermediate scale in the SUSY spectrum.

\subsection{SUSY $SO(10)$ models with a sliding $SU(2)_{R}$~scale 
\label{SU2Rmodels}}
\subsubsection{Models~I and II: single sliding intermediate scale }

First, we shall introduce two variants of the model advocated in
\cite{Dev:2009aw} which supply the original setting with a few extra
ingredients in order to make it potentially realistic, c.f.,
Sect.~\ref{sect:models:generalities}.
\vskip 3mm
\paragraph{Model~I:\label{ModelI}}
The field content relevant to the running in this Model~is specified
in TABLE~\ref{tab:ModelI}. 

\begin{table}[t,floatfix]
\begin{tabular}{|c|c|c|c|}
\hline
{Field} & Multiplicity & $3_{c}2_{L}2_{R}1_{B-L}$& $SO(10)$ origin\\
\hline
$Q$ & 3 & $(3,2,1,+\tfrac{1}{3})$ & 16\\
$Q^{c}$ & 3 & $(\bar 3,1,2,-\tfrac{1}{3})$ & 16\\
$L$  & 3 & $(1,2,1,-1)$& 16\\
$L^{c}$& 3 & $(1,1,2,+1)$  & 16\\
$S$ &  3 & $(1,1,1,0)$ & 1\\
$\delta_{d}$, $\bar\delta_{d}$ & 1 & $(3,1,1,- \tfrac{2}{3})$, $(\bar 3,1,1,+\tfrac{2}{3})$ 
& 10\\
\hline
$\Phi$ & 1 & $(1,2,2,0)$ & $10$, $120$\\
$\chi$, $\overline\chi$ & 1 & $(1,2,1,\pm 1)$ 
 & $\overline{16}$, $16 $\\
$\chi^{c}$, $\overline\chi^{c}$ & 3 & $(1,1,2,\mp 1)$ 
 & $\overline{16}$, $16 $\\
\hline 
\end{tabular}
\caption{The relevant part of the field content of Model~I with a
sliding $SU(2)_{R}$-breaking scale discussed in section
\ref{ModelI}. In the third column the relevant fields are
characterized by their $SU(3)_{c}\times SU(2)_{L}\times
SU(2)_{R}\times U(1)_{B-L}$ quantum numbers while their $SO(10)$
origin is specified in the fourth column. \label{tab:ModelI}}
\end{table}

The original $SO(10)$ gauge symmetry is broken down to the MSSM in two
steps via an intermediate $SU(3)_{c}\times SU(2)_{L}\times
SU(2)_{R}\times U(1)_{B-L}$ symmetry stage. The first step, i.e., the
$SO(10)\to SU(3)_{c}\times SU(2)_{L}\times SU(2)_{R}\times
U(1)_{B-L}$ breaking, is triggered by an interplay between the VEVs of
$45$ and $210$ of $SO(10)$. Subsequently, the $SU(2)_{R}\times
U(1)_{B-L}$ gauge symmetry is broken down to the $U(1)_{Y}$ of
hypercharge by the VEVs of the $\chi^{c}\oplus \overline{\chi}^{c}$
pair which can emerge for instance from its cubic interaction with a
full singlet $\rho$\footnote{It is perhaps worth mentioning that, for
a very low scale of the $SU(2)_{R}\times U(1)_{B-L}$ breaking, the
relevant VEV can be devised even without an extra singlet because,
then, the interplay between a ``RH $\mu$-term'' for $\chi^{c}\oplus
\overline{\chi}^{c}$ and the relevant soft mass should be sufficient,
in complete analogy with the $SU(2)_{L}$-doublet sector in the
MSSM.}. Note that, at the one-loop level, such a neutral $\rho$ field
can be put essentially anywhere between $M_{GUT}$ and $M_{SUSY}$
without any impact on the gauge unification.  The $b_{i}$-coefficients
at the $SU(3)_{c}\times SU(2)_{L}\times SU(2)_{R}\times
U(1)_{B-L}$ level read $b_{3}=-2$, $b_{L}=2$, $b_{R}=4$ and
$b_{B-L}^{\rm c}=13$ where the last number corresponds to the
canonically normalized $B-L$ charge, which is obtained from the
``physical'' one (based on $B^{\rm p}_{Q}=+\tfrac{1}{3}$ and $L^{\rm
p}_{L}=+1$) by means of the formula $(B-L)^{\rm
c}=\sqrt{\tfrac{3}{8}}\,(B-L)^{\rm p}$. Note that these coefficients
happen to be entirely identical to the setting advocated
in \cite{Dev:2009aw} and, thus, the leading-log evolution of the soft
SUSY-breaking parameters in that Model~is covered by our analysis of
Model~I, see Sect.~\ref{SU2RmodelsResults}. The scale of
the $SU(2)_{R}\times U(1)_{B-L}$ breaking is not determined because
it drops from the formula for the unification scale (owing namely to
the hypercharge-matching condition
$\alpha^{-1}_{Y}=\tfrac{3}{5}\alpha^{-1}_{R}+\tfrac{2}{5}\alpha^{-1}_{B-L}$)
and affects only the value of the GUT-scale gauge coupling
$\alpha_{G}$ which, however, is subject of much weaker constraints.

\begin{figure}[t]
\parbox{8cm}{
\includegraphics[width=8cm]{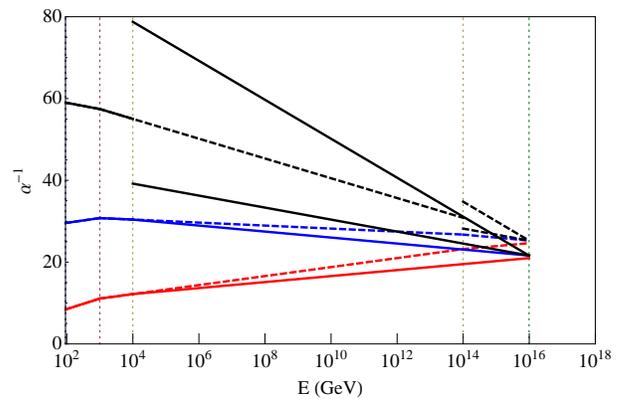}
}
\caption{Gauge coupling unification in Model~I in two limits
corresponding to different positions of the sliding $SU(2)_{R}\times
U(1)_{B-L}$ breaking scale $V_{R}$. In solid lines, we depict the RGE
behaviour of the gauge couplings for $V_{R}$ in the vicinity of the
electroweak scale $V_{R}\sim 10^{4}$GeV while the dashed lines
correspond to $V_{R}\sim 10^{14}$GeV. The position of the intersection
region shifts slightly up with rising $V_{R}$ but the corresponding
scale remains intact.\label{fig:ModelIrunning}}
\end{figure}

This, on one hand, makes the gauge coupling unification in Model~I
qualitatively similar to the MSSM case, see FIG.~\ref{fig:ModelIrunning}.
On the other hand, it is well known that in the MSSM the one-loop
gauge unification is incompatible with the latest extractions of
$\alpha_{s}$ unless the soft SUSY-breaking scale is pushed well below
1 TeV. This few percent mismatch is expected to be accounted for by
GUT-scale thresholds whose detailed analysis is, however, beyond the
scope of this work. Thus, in what follows, we shall simply parametrize
our ignorance of the shape of the GUT spectrum by considering
unification regions from where the $SU(3)_{c}\times SU(2)_{L}\times
SU(2)_{R}\times U(1)_{B-L}$ gauge couplings can emanate rather than
unique unification points, c.f., FIG.~\ref{fig:triangle} and
discussion in Sect.~\ref{SU2RmodelsResults}. Though this approach is 
oversimplified in several aspects, it admits to estimate the
magnitude of the theoretical error associated to the lack of exact
gauge-coupling unification in this model.

Concerning the effective flavour structure of the model, there are two
aspects worth some discussion here, namely, the structure of the
effective MSSM Higgs doublet pair $\Phi$ and the possibility to
accommodate the SM quark and lepton masses and mixing (requirements 3
and 4 formulated at the beginning of this Section). First, the
effective L-R bidoublet $(1,2,2,0)$ corresponds to a massless
combination of the $(1,2,2)$ and $(15,2,2)$ Pati-Salam components of
$10$ and $120$ of $SO(10)$, respectively, which can mix at the
GUT-level due to the PS-breaking VEV in an $SO(10)$-breaking multiplet
such as $45$ and/or $210$. Usually, the role of the extra Higgs such
as 120 and/or 126 in the Yukawa sector is namely to provide
Clebsch-Gordan coefficients that would break the degeneracy of the
effective Yukawa couplings among up and down quarks and charged
leptons.  However, an extra $120$ alone is still not enough as it does
not yield enough freedom to accommodate the SM data
\cite{Lavoura:2006dv}. Actually, the issue becomes even worse if the
MSSM-level mass matrices for the two hypercharge components of $\Phi$
are virtually identical, as one can expect for a single bidoublet at
play in the low-scale $SU(2)_{R}$-breaking regime. Both issues are
potentially resolved due to the extra vector-like down-type quark pair
$\delta_{d}\oplus \delta_{d}$ and an additional $SU(2)_{L}$-doublet
Higgs pair $\chi\oplus \overline{\chi}$ (which, simultaneously, ensure
the right $b$-coefficients for the running), c.f.,
TABLE~\ref{tab:ModelI}. In this case, the down-type quark mass matrix
is extended\footnote{For an explicit $SO(10)$ realisation of this
mechanism see e.g. \cite{Heinze:2010du} and references therein.} to
$4\times 4$ which, together with the extra freedom in the MSSM doublet
sector, should be enough to avoid the Grimus-Kuhbock-Lavoura (GKL)
no-go \cite{Lavoura:2006dv}. Let us also mention that the VEV of
$\chi_{L}$ gives rise to the $LS$ entry in the neutrino mass matrix
generating the linear seesaw mechanism and, unlike in
\cite{Malinsky:2005bi}, it is not naturally suppressed in this case
because $\chi\oplus \overline{\chi}$ resides well below the GUT
scale. Thus, one has to assume a small $LS\chi$ Yukawa coupling.

\paragraph{Model~II:\label{ModelII}}
\begin{table}[t]
\begin{tabular}{|c|c|c|c|}
\hline
{Field} & Multiplicity & $3_{c}2_{L}2_{R}1_{B-L}$& $SO(10)$ origin\\
\hline
$Q$ & 3 & $(3,2,1,+\tfrac{1}{3})$ & 16\\
$Q^{c}$ & 3 & $(\bar 3,1,2,-\tfrac{1}{3})$ & 16\\
$L$& 3  & $(1,2,1,-1)$ & 16\\
$L^{c}$ & 3 & $(1,1,2,+1)$ & 16\\
$S$ & 3 & $(1,1,1,0)$ & 1\\
$\delta_{u}$, $\bar\delta_{u}$ & 1 & $(3,1,1, +\tfrac{4}{3})$, $(\bar 3,1,1,-\tfrac{4}{3})$ 
& 45\\
\hline
$\Phi$ & 1 & $(1,2,2,0)$ & $10$, $120$\\
$\chi$, $\overline\chi$ & 1& $(1,2,1,\pm 1)$ 
 & $\overline{16}$, $16 $\\
$\chi^{c}$, $\overline\chi^{c}$ & 2  & $(1,1,2,\mp 1)$ 
& $\overline{16}$, $16 $\\
\hline 
\end{tabular}
\caption{The same as in TABLE \ref{tab:ModelI} for Model~II defined in
Sect.~\ref{ModelII}. The main variation with respect to Model~I is the
$B-L$ charge of the vector-like colour triplet pair owing to its
different $SO(10)$ origin. The extra $\delta_{u}$ and
$\overline{\delta}_{u}$ fields can mix with the up-type quarks at the
MSSM level which leads to a potentially realistic effective flavour
structure. In order to maintain the MSSM-like unification pattern, the
number of the $SU(2)_{R}$ doublets has been reduced, thus making the
setting slightly more compact than in Model~I.\label{tab:ModelII}}
\end{table}

\begin{figure}[t]
\parbox{8cm}{
\includegraphics[width=8cm]{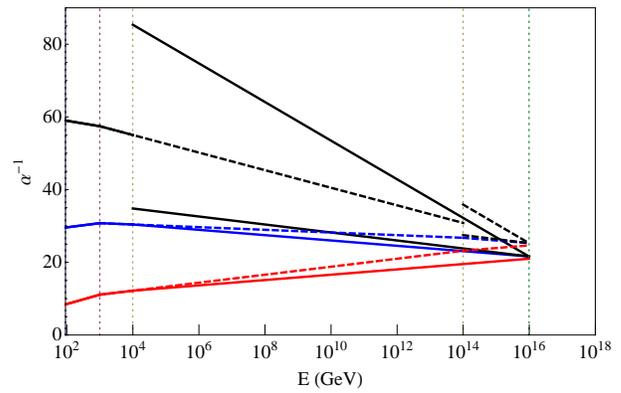}
}
\caption{Gauge coupling unification in Model~II in two limits
corresponding to different positions of the sliding $SU(2)_{R}\times
U(1)_{B-L}$ breaking scale $V_{R}$. In solid lines, we depict the RGE
behaviour of the gauge couplings for $V_{R}$ in the vicinity of the
electroweak scale $V_{R}\sim 10^{4}$GeV while the dashed lines
correspond to $V_{R}\sim 10^{14}$GeV. The position of the intersection
region shifts slightly up with rising $V_{R}$ but the corresponding
scale remains intact.\label{fig:ModelIIrunning}}
\end{figure}

The relevant $b_{i}$-coefficients at the $SU(3)_{c}\times
SU(2)_{L}\times SU(2)_{R}\times U(1)_{B-L}$ level read $b_{3}=-2$,
$b_{L}=2$, $b_{R}=3$ and $b_{B-L}^{\rm can}=29/2$. Indeed, these
numbers differ from Model~I only in the $SU(2)_{R}\times U(1)_{B-L}$
sector and the variations in the relevant $b$-coefficients obey
$\Delta b_{R}+\tfrac{2}{3}\Delta b_{B-L}=0$ so the $b$-coefficient
associated to the ``effective'' MSSM hypercharge is the same as in
Model~I. Therefore, apart from the difference in the specific slopes
of the $SU(2)_{R}\times U(1)_{B-L}$ curves the qualitative picture of
the gauge coupling unification in Model~II,
c.f. FIG.~\ref{fig:ModelIIrunning}, is very similar to that observed
in Model~I.
Nevertheless, as we shall see in Sect.~\ref{SU2RmodelsResults}, even
such a slight change in the gauge-coupling behaviour at the
$SU(3)_{c}\times SU(2)_{L}\times SU(2)_{R}\times U(1)_{B-L}$ level
is enough to generate a significant difference between the Model-I and
Model-II soft invariants, especially if the $SU(3)_{c}\times
SU(2)_{L}\times SU(2)_{R}\times U(1)_{B-L}$ running is
long. However, if the $SU(2)_{R}\times U(1)_{B-L}$ gauge symmetry
happens to be broken close to the GUT scale, the two models will be
indistinguishable from the soft-sector point of view.

Concerning the flavour structure of Model~II, it is indeed very
similar to that of Model~I, with the main difference that here the GKL
no-go \cite{Lavoura:2006dv} is overcome by a $4\times 4$ extension of
the up-type quark mass matrix. Moreover, since it is the VEV of
$\overline{\chi}$ rather than that of $\chi$ that enters the extended
up-type quark mass matrix, $\langle\chi\rangle$ can be made much
smaller than $\langle\overline{\chi}\rangle$ which also relieves the
need for the small $LS\chi$ Yukawa in the neutrino sector. Given also
the reduced number of the $SU(2)_{R}$ doublets, Model~II constitutes a
somewhat more compact alternative to Model~I.

%
\subsubsection{Model~III: sliding $SU(2)_{R}$ and Pati-Salam scales
\label{ModelIII}}

The third model of our interest belongs to the second category of the
simple classification given in
Sect.~\ref{sect:models:generalities}. In particular, the sliding
nature of the $SU(2)_{R}\times U(1)_{B-L}$ scale is achieved via an
interplay with another intermediate scale, namely, the Pati-Salam
$SU(4)_{C}\times SU(2)_{L}\times SU(2)_{R}$, rather than a delicate
adjustment \`a la Model~I or Model~II owing to  very specific field
contents. Thus, the initial $SO(10)$ gauge symmetry is broken down to
the MSSM in three steps. The field content relevant to the two
intermediate-symmetry stages is given in TABLE~\ref{tab:ModelIII}.
\begin{table}[t]
\begin{tabular}{|c|c|c|c|c|c|}
\hline
{Field} & Mult. & $3_{c}2_{L}2_{R}1_{B-L}$ & Pati-Salam & $SO(10)$\\
\hline
$Q$ & 3 & $(3,2,1,+\tfrac{1}{3})$ & $(4,2,1)$ & 16\\
$Q^{c}$ & 3 & $(\bar 3,1,2,-\tfrac{1}{3})$ & $(\bar 4,1,2)$ & 16\\
$L$ &  3 & $(1,2,1,-1)$ & $(4,2,1)$& 16\\
$L^{c}$& 3  & $(1,1,2,+1)$ & $(\bar 4,1,2)$ & 16\\
$\Sigma^{c}$& 3  & $(1,1,3,0)$ & $(1,1,3)$  & 45\\
$\delta_{d}$, $\bar\delta_{d}$& 1  & $(3,1,1,\mp \tfrac{2}{3})$
& $(6,1,1)$ & 10\\
\hline
$\Phi$& 2  & $(1,2,2,0)$ & $(1,2,2)$ & $10$\\
$\Omega$ & 1 & $(1,1,3,0)$ & $(1,1,3)$ & $45$\\
$\chi$, $\overline\chi$& 1  & $(1,2,1,\pm 1)$
& $(\bar 4,2,1)$,$(4,2,1)$ & $\overline{16}$, $16 $\\
$\chi^{c}$, $\overline\chi^{c}$& 1 & $(1,1,2,\mp 1)$
& $ (4,1,2)$,$(\bar 4,1,2)$ & $\overline{16}$, $16 $\\
$\Psi$ & 1 & absent & $(15,1,1)$ & $45$\\
\hline 
\end{tabular}
\caption{The effective field contents of Model~III in the two
intermediate symmetry stages. \label{tab:ModelIII}}
\end{table}
In more detail, the initial $SO(10)\to SU(4)_{C}\times
SU(2)_{L}\times SU(2)_{R}$ breaking is triggered by the GUT-scale VEV
of $54$ of $SO(10)$. The subsequent $SU(4)_{C}\times SU(2)_{L}\times
SU(2)_{R}\to SU(3)_{c}\times SU(2)_{L}\times SU(2)_{R}\times
U(1)_{B-L}$ breaking is due to the VEV of the $\Psi$ emerging again
from its interplay with an extra singlet. Finally, the
$SU(2)_{R}\times U(1)_{B-L}$ symmetry is broken down to the MSSM by
means of the VEVs of $\chi^{c}\oplus \overline{\chi}^{c}$ which are
connected by the $B-L$-neutral $SU(2)_{R}$-triplet $\Omega$.  At the
Pati-Salam stage, the $b_{i}$-coefficients read $b_{4}=3$, $b_{L}=6$,
$b_{R}=14$ while at the $SU(3)_{c}\times SU(2)_{L}\times
SU(2)_{R}\times U(1)_{B-L}$ level they are $b_{3}=-2$, $b_{L}=3$,
$b_{R}=11$ and $b_{B-L}^{\rm can}=10$.

\begin{figure}[t]
\parbox{8cm}{
\includegraphics[width=8cm]{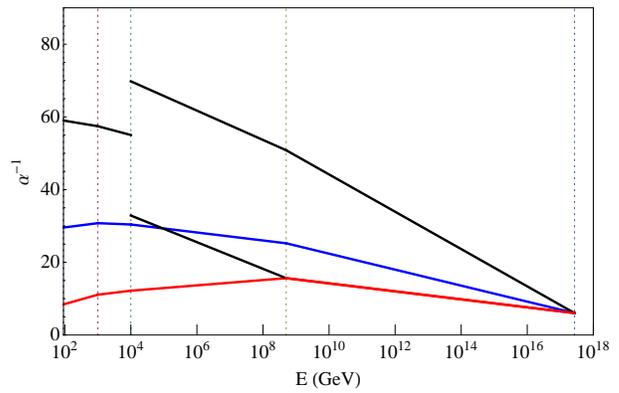}
}
\caption{Running in the Model~III variant of the low-LR scale SUSY
SO(10). Here the SO(10) gauge symmetry is broken first into a
Pati-Salam intermediate stage stretching from the unification point
down to the relevant energy scale $V_{PS}$ (in the middle) and,
subsequently, to the L-R symmetry stage. The value of $V_{PS}$ is
correlated to the position of the L-R breaking scale $V_{R}$ which can
again slide from as low as few TeV up to roughly $10^{14}$GeV, c.f.,
FIG.~\ref{fig:parametricspacePS}.\label{fig:ModelIIIrunning}}
\end{figure}

In this model, both the position of the GUT scale as well as the value
of $\alpha_{G}$ depend on both intermediate scales. However, unlike in
Models~I and II, here the gauge unification can always be made exact,
c.f., FIG.~\ref{fig:ModelIIIrunning}, even at the one-loop level, and,
thus, there is no extra theoretical uncertainty other than the error
in the electroweak-scale $\alpha_{s}$ to be taken into account.

The flavour structure of this model relies on the presence of three
extra copies of $SU(2)_{R}$ triplet $\Sigma^{c}$ which in the neutrino
sector play a role similar to that of $S$ in Models~I and II. In
particular, they expand the $6\times 6$ neutrino mass matrix to
$9\times 9$ where, e.g., the $L^{c}\Sigma^{c}$ sector comes from the
contraction with the VEV of $\chi^{c}$, but without any entry
generated at the $L\Sigma^{c}$ ``linear seesaw'' position. Thus, there
is no need for an extra fine-tuning in the seesaw formula in Model
III.  Moreover, the charged components of $\Sigma^{c}$ can mix with the
charged leptons and, hence, provide the welcome departure from the
down-type quarks even if the MSSM doublets span over 10's of $SO(10)$
only. Indeed, the relevant $6\times 6$ charged-lepton mass matrix
looks schematically like \be M_{\ell}\propto \left(\begin{array}{cc}
Yv^{10}_{d} & 0 \\ \langle \overline\chi^{c}\rangle &
\mu_{\Sigma^{c}}+\langle \Omega\rangle
\end{array}\right)\,,
\ee where the row and column bases are $\{L, \Sigma^{c-}\}$ and
$\{L^{c}, \Sigma^{c+}\}$, respectively, and $\mu_{\Sigma^{c}}$ is the
associated singlet mass parameter. Note also that the VEV of $\Omega$
is antisymmetric in the generation space and, thus, does not
contribute to the neutrino Majorana mass matrix. Finally, the two MSSM
Higgs doublets are different because the underlying bi-doublets
contract through $\Omega$ and, therefore, the effective up-type quark
Yukawa coupling differs from the down-type one even without the need
to resort to the mixing with the vector-like
$\delta_{d}\oplus\bar\delta_{d}$ pair.


\subsection{SUSY $SO(10)$ models with a sliding $U(1)_{R}$ scale\label{ModelIV}}
All the models discussed so far featured an intermediate
$SU(2)_{R}\times U(1)_{B-L}$ symmetry which, at a certain scale, was
broken directly down to the $U(1)_{Y}$ of the MSSM hypercharge. The
full $SU(2)_{R}$, however, is not the minimal option to realize a
gauge symmetry acting in the RH sector of the matter spectrum in a way
compatible with the MSSM quantum numbers. Indeed, the hypercharge
sum-rule $Y=T_{R}^{3}+(B-L)/2$ trivially holds even if one sticks to
the $U(1)_{R}$ subgroup of the original $SU(2)_{R}$ generated by
$T_{R}^{3}$ alone.

On the other hand, within $SO(10)$ broken down to an intermediate
$SU(3)_{c}\times SU(2)_{L}\times U(1)_{R}\times U(1)_{B-L}$ stage,
only $Z'$ and the associated $U(1)_{R}\times U(1)_{B-L}\to
U(1)_{Y}$-breaking Higgs fields remain light (at least in minimally
fine-tuned scenarios) and, thus, the intermediate-scale dynamics is
generally much simpler than in the models based on the full
$SU(2)_{R}\times U(1)_{B-L}$. In view of that, one can expect that
also the intermediate-scale dependence of the soft RGE invariants will
be much milder than in the former case. Moreover, with more than a
single abelian gauge factor at play, there is a new class of effects
associated with the so called kinetic mixing between the associated
gauge fields. Both these aspects make this class of models worth 
further scrutiny.

\subsubsection{General remarks}
Remarkably, the simplicity of the minimally fine-tuned
$U(1)_{R}\times U(1)_{B-L}\to U(1)_{Y}$ scenarios automatically
implies the scale of this spontaneous symmetry breakdown is a sliding
one. Indeed, minimal fine-tuning implies that the spectrum of the
Model~in the unbroken phase consists of that of the MSSM plus $Z'$
plus an MSSM-neutral Higgs responsible for the relevant symmetry
breaking. Since the gauge field associated to the hypercharge
($B_{Y}$) does not feel any effect of either $Z'$ nor the
hypercharge-neutral Higgs\footnote{To put this statement on a firm
ground the effects of the kinetic mixing must be considered, see,
e.g., \cite{Fonseca:2011vn}.} the ``effective'' hypercharge gauge
coupling (corresponding to a relevant combination of $g_{Y}$ and
$g_{R}$) in this picture runs as if it were in the MSSM, at least at
the one-loop level. Thus, the specific position of the
$U(1)_{R}\times U(1)_{B-L}\to U(1)_{Y}$ breaking scale is, in this
case, irrelevant for the one-loop gauge running.

This, however, is not the case for the leading-log soft RGE invariants
of our interest. In particular, unlike $B_{Y}$, both the
$U(1)_{R}\times U(1)_{B-L}$ gauge bosons enter the renormalized
propagators of squarks and sleptons \be
\parbox{8.5cm}{
\includegraphics[width=4cm]{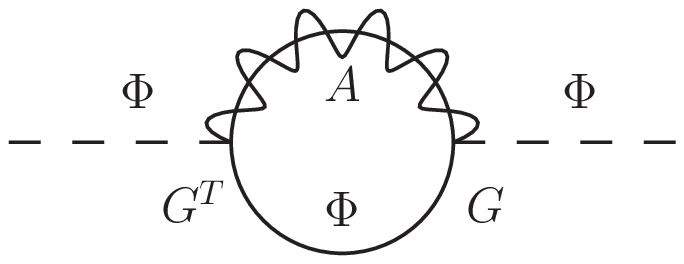}\includegraphics[width=4cm]{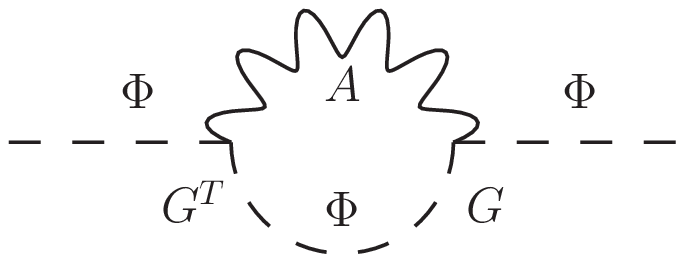}
} \ee and one can expect a residual dependence of the invariants on
the $U(1)_{R}\times U(1)_{B-L}$-breaking scale. Nevertheless, as we
shall demonstrate in a particular realization of this simple scheme,
such effects should be much milder than those in the scenarios with
the full gauged $SU(2)_{R}$ symmetry.

\subsubsection{Model~IV: $U(1)_{R}\times U(1)_{B-L}\to U(1)_{Y}$ breaking}
Here we consider a variant of the basic SUSY $SO(10)$ model advocated
in \cite{Malinsky:2005bi} in which an extended intermediate
$U(1)_{R}\times U(1)_{B-L}$ stage follows a short $SU(2)_{R}\times
U(1)_{B-L}$ phase. The field content relevant to the RG running in
the first two parts of the symmetry-breaking chain is given in
TABLE~\ref{tab:ModelIV}.
\begin{table}[t]
\begin{tabular}{|c|c|c|c|c|}
\hline
{Field} & Mult. & $3_{c}2_{L}1_{R}1_{B-L}$ & $3_{c}2_{L}2_{R}1_{B-L}$& $SO(10)$ \\
\hline
$Q$ & 3 &$(3,2,0,+\tfrac{1}{3})$ &$(3,2,1,+\tfrac{1}{3})$ &  $16$\\
$Q^{c}$ & 3 & $(\bar 3,1,\pm \tfrac{1}{2},-\tfrac{1}{3})$
& $(\bar 3,1,2,-\tfrac{1}{3})$ &  16\\
$L$ & 3 & $(1,2,0,-1)$ & $(1,2,1,-1)$ &  $16$ \\
$L^{c}$ & 3 & $(1,1,\pm \tfrac{1}{2},+1)$
& $(1,1,2,+1)$ &  16\\
$S$ & 3 & $(1,1,0,0)$ & $(1,1,1,0)$ &  1\\
\hline
$\Phi$ & 2 & $(1,2,\pm\tfrac{1}{2},0)$
& $(1,2,2,0)$ &  10\\
$\Omega$ & 1 & absent & $(1,1,3,0)$ &  45\\
$\chi$, $\overline\chi$ & 1 & absent & $(1,2,1,\pm 1)$ 
& $\overline{16}$, $16 $\\
$\chi^{c}$, $\overline\chi^{c}$ & 1 & $(1,1,\pm \tfrac{1}{2},\mp 1)$
& $(1,1,2,\mp 1)$
&  $\overline{16}$, $16 $\\
\hline 
\end{tabular}
\caption{The effective field contents of Model~IV relevant to the two
intermediate symmetry stages.\label{tab:ModelIV}}
\end{table}

In more detail, after the initial $SO(10)\to SU(3)_{c}\times
SU(2)_{L}\times SU(2)_{R}\times U(1)_{B-L}$ breaking triggered by
essentially the same mechanism as in Models~I and II, the subsequent
$SU(2)_{R}\times U(1)_{B-L}\to U(1)_{R}\times U(1)_{B-L}$ requires a
VEV of the $\Omega$ field which, at the level of an effective theory,
can again emerge from its interplay with a LR singlet. The last
symmetry-breaking step is then achieved in a similar manner by the
VEVs of the MSSM-neutral components of the
$\chi^{c}\oplus\overline{\chi}^{c}$ fields.

\begin{figure}[t]
\parbox{8cm}{
\includegraphics[width=8cm]{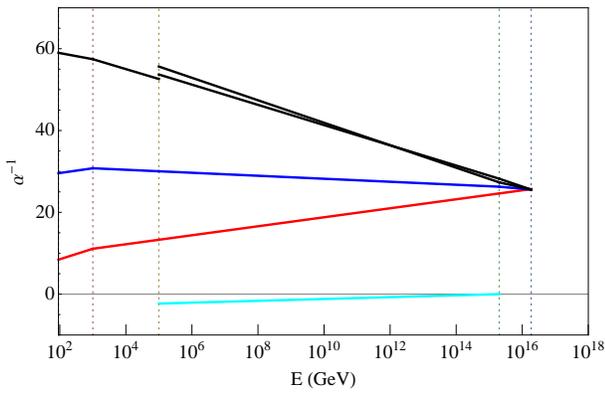}
}
\caption{Running in Model~IV with a low $B-L$ scale. Note the effects
of the U(1) mixing in the running \& matching; the lowest curve 
corresponds to the off-diagonals of the $(GG^{T}/4\pi)^{-1}$ matrix.
\label{fig:ModelIVrunning}}
\end{figure}

The relevant $b_{i}$-coefficients at the $SU(3)_{c}\times
SU(2)_{L}\times SU(2)_{R}\times U(1)_{B-L}$ level read $b_{3}=-3$,
$b_{L}=2$, $b_{R}=5$ and $b_{B-L}^{\rm can}=15/2$.  In the
$SU(3)_{c}\times SU(2)_{L}\times U(1)_{R}\times U(1)_{B-L}$ stage,
however, the effects of the $U(1)$ mixing must be taken into account
and, thus, the $b$-coefficients in the $U(1)_{R}\times U(1)_{B-L}$
sector constitute a matrix of anomalous dimensions $\gamma$. One has
$b_{3}=-3$, $b_{L}=1$ and \be \gamma^{\rm
phys}=\left(\begin{array}{cc} 15/2 & -1\\ -1 & 18
\end{array}\right)\,,
\ee which should be brought into the canonical basis by means of a
normalization matrix $N=\text{diag}(1,\sqrt{3/8})$, $\gamma^{\rm
can}=N\gamma^{\rm phys}N$.  The details of the one-loop RGE evolution
of gauge couplings and soft masses in theories with more than a single
abelian gauge factor are summarized in Appendix~\ref{app:mixedrunning}. 
The qualitative features of the gauge-coupling running in this setting
can be seen in FIG.~\ref{fig:ModelIVrunning}.

Concerning the flavour structure of the model, the situation is
essentially identical to that described in the original work
\cite{Malinsky:2005bi}. The only exception is a second LR bi-doublet
retained until the $U(1)_{R}\times U(1)_{B-L}$ breaking scale which
might be necessary in order to get a potentially realistic pattern of
the effective Yukawa couplings. Nevertheless, the salient features of
the Model~in the soft sector should not depend much on the detailed
realization of the effective Yukawa pattern.

\section{Leading-log RGE invariants\label{sect:invariants}}

In this section we focus on the calculation of the invariants 
using mSugra boundary conditions. mSugra is defined 
at the GUT-scale, $M_{\rm G}$, by a common gaugino mass $M_{1/2}$, 
a common scalar mass $m_0$ and the trilinear coupling $A_0$, which 
gets multiplied by the corresponding Yukawa couplings to obtain the 
trilinear couplings in the soft SUSY breaking Lagrangian 
\footnote{It is sometimes argued that this setup should better be 
called CMSSM, since there are even simpler models of supergravity 
type breaking in which $A_0$ is not a free parameter, as for example 
in Polonyi type supergravity  \cite{Ellis:2003pz,Martin:1997ns}. 
Since we  will be concerned with only the first two sfermion 
generations this distinction is irrelevant for us.}.
In addition, at the electro-weak scale, $\tan\beta =v_u/v_d$ is fixed. 
Here, as usual, $v_d$ and $v_u$ are the vacuum expectation values 
(vevs) of the neutral components of $H_d$ and $H_u$, respectively.  
Finally, the sign of the $\mu$ parameter has to be chosen.

Renormalization group equations for general supersymmetric models 
are known up to 2-loop order \cite{Martin:1993zk}. The only case not 
covered in the otherwise general equations given in \cite{Martin:1993zk} 
are supersymmetric models with more than one $U(1)$ group. 
With more than a single abelian gauge factor, there appears a new 
class of effects associated with the so called kinetic mixing 
between the associated gauge fields. RGEs for this case have been 
derived very recently in \cite{Fonseca:2011vn}.

Barring for the moment the effects of $U(1)$ mixing present in Model IV, 
at the 1-loop level, one can devise a simple set of analytic equations 
for the soft terms. Gaugino masses scale as gauge couplings 
do and so the requirement of GCU fixes the gaugino masses at the 
low scale 
\begin{eqnarray}
M_i(m_{SUSY}) = \frac{\alpha_i(m_{SUSY})}{\alpha(M_G)} M_{1/2}.
\label{eq:gaugino}
\end{eqnarray}
Eq. (\ref{eq:gaugino}) implies that the relationship of the $M_i$ 
to $M_{1/2}$ is changed in Models I to III, since $\alpha(M_G)$ 
is shifted. 

Neglecting the Yukawa couplings for the soft mass parameters of the 
first two generations of sfermions one can write
\begin{eqnarray}\label{eq:scalar}
m_{\tilde f}^2  &=& m_0^2  +  \frac{M_{1/2}}{\alpha(M_{\rm G})^2}
\sum_{R_j} \sum_{i=1}^N  {\tilde f}_i^R \alpha_i(v_{R_j})^2\,.
\end{eqnarray}
Here, the sum over ``$R_j$'' runs over the different regimes in 
the models under consideration, while the sum over $i$ runs 
over all gauge groups in a given regime. $\alpha_i(v_{R_j})$ is 
to be understood as the gauge coupling of group $i$ evaluated 
at the upper end of regime $R_j$. In the MSSM one would 
have only to consider one regime, namely from the SUSY scale to 
the GUT scale. In Models I and II we have two different regimes, 
while in Models III and IV there are a total of three regimes 
to consider. 

The different ${\tilde f}_i^R$ can be written in a compact form
as:
\begin{eqnarray}\label{eq:fi}
 \tilde{f}^{R}_i &=& \frac{c_i^{f,R}}{b_i} \left[
1 - \left(\frac{\alpha_i(v_x)}{\alpha_i(v_y)}\right)^2\right]\,,
\end{eqnarray}
where $v_x$ and  $v_y$, respectively, indicate the value of the relevant $\alpha$ at the lower and higher 
boundaries of the regime under consideration. 
The $c^{f,R}_i$ coefficients given in TABLE~\ref{tab:coeff} are
proportional to the values of the quadratic Casimir of representation
$R_{f}$ hosting the matter field $f$ with respect to the group $G$ in the regime~$R$
\begin{equation}
c^{f,R}_i=2C_{G}(R_{f})\,.
\end{equation}
They are readily evaluated from the basic formula 
\be
C_{G}(R)d(R)=T_{2}(R)d(G)\,,  
\ee 
where $d(G)$ is the dimension of the
group $G$, $T_2(R)$ the Dynkin index of the representation $R$ and
$d(R)$ is its dimension. Note that the coefficients $c^{f,R}_i$ are different
for the different fermions, which leads to a different coefficient in
front of $M_{1/2}$ in eq. (\ref{eq:scalar}). The $b_i$ in eq.
(\ref{eq:fi}) are the one-loop $b$-coefficients for the different models
defined in the previous section. For completeness, the well-known one-loop beta-coefficients for the MSSM are (in the traditional $SU(5)$
normalization): $b=(b_1,b_2,b_3)^{MSSM}= \textstyle
(\frac{33}{5},1,-3)$.

Eq. (\ref{eq:scalar}) is valid neglecting $U(1)$-mixing effects.  The
extra effects due to the kinetic $U(1)$ mixing relevant in Model IV
are summarized in the Appendix~\ref{app:mixedrunning}; for a more
detailed discussion including higher-loop effects, see
\cite{Fonseca:2011vn}.

\begin{table}[ht]
\begin{tabular}{|c|ccccc|}
\hline
$\tilde f$ & $\tilde E$ & $\tilde L$ &$\tilde D$ & $\tilde U$ & $\tilde Q$ \\
\hline
MSSM & & & & & \\
\hline
$c^{f,MSSM}_1$ & $\frac{6}{5}$ & $\frac{3}{10}$ & $\frac{2}{15}$
                 & $\frac{8}{15}$ & $\frac{1}{30}$ \\
$c^{f,MSSM}_2$ & 0 &  $\frac{3}{2}$ &  0 & 0 & $\frac{3}{2}$ \\
$c^{f,MSSM}_3$ & 0 & 0 &  $\frac{8}{3}$ & $\frac{8}{3}$ & $\frac{8}{3}$ 
\\ \hline
\hline
$U(1)_R\times U(1)_{B-L}$ & & & & & \\
\hline
$c^{f,BL}_{BL}$ & $\frac{3}{4}$ & $\frac{3}{4}$ & $\frac{1}{12}$
                 & $\frac{1}{12}$ & $\frac{1}{12}$ \\
$c^{f,BL}_L$ & 0 &  $\frac{3}{2}$ &  0 & 0 & $\frac{3}{2}$ \\
$c^{f,BL}_R$ & $\frac{1}{2}$ & 0 & $\frac{1}{2}$ & $\frac{1}{2}$ & 0 \\
$c^{f,BL}_3$ & 0 & 0 &  $\frac{8}{3}$ & $\frac{8}{3}$ & $\frac{8}{3}$ 
\\ \hline
\hline
LR & & & & & \\
\hline
$c^{f,LR}_{BL}$ & $\frac{3}{4}$ & $\frac{3}{4}$ & $\frac{1}{12}$
                 & $\frac{1}{12}$ & $\frac{1}{12}$ \\
$c^{f,LR}_L$ & 0 & $\frac{3}{2}$ & 0  & 0 & $\frac{3}{2}$ \\
$c^{f,LR}_R$ & $\frac{3}{2}$ & 0 &$\frac{3}{2}$ 
&$\frac{3}{2}$ & 0 \\
$c^{f,LR}_3$ & 0 & 0 &  $\frac{8}{3}$ & $\frac{8}{3}$ & $\frac{8}{3}$ 
\\ \hline
Pati-Salam & & & & & \\
\hline
$c^{f,PS}_L$ & 0 & $\frac{3}{2}$ & 0 & 0 & $\frac{3}{2}$ \\ 
$c^{f,PS}_R$ & $\frac{3}{2}$ & 0 & $\frac{3}{2}$ &$\frac{3}{2}$ & 0 \\
$c^{f,PS}_4$ & $\frac{15}{4}$ &$\frac{15}{4}$ & $\frac{15}{4}$ &
$\frac{15}{4}$ &$\frac{15}{4}$ 
\\ \hline
\end{tabular}
\caption{Coefficients $c^{\tilde f}_i$ for eq.~(\ref{eq:scalar}) for
different symmetry stages. The MSSM and the LR parts are relevant to
all four models under consideration; the $U(1)_R\times U(1)_{B-L}$ and
the Pati-Salam parts are used solely for Model IV and Model III,
respectively.}
\label{tab:coeff}
\end{table}

Individual SUSY masses depend strongly on the initial values for 
$m_0$ and $M_{1/2}$. However, one can form four different combinations, 
which we choose to be
\begin{eqnarray}\label{eq:4Inv}
LE \equiv (m_{\tilde L}^2 -m_{\tilde E}^2)/M_1^2, \\ \nonumber
QE \equiv (m_{\tilde Q}^2 -m_{\tilde E}^2)/M_1^2, \\ \nonumber
DL \equiv (m_{\tilde D}^2 -m_{\tilde L}^2)/M_1^2,\\ \nonumber
QU \equiv (m_{\tilde Q}^2 -m_{\tilde U}^2)/M_1^2. 
\end{eqnarray}
It is easy to see that, at the leading-log level, $m_0$ and $M_{1/2}$ 
drop out of the equations for the invariants. Note, that 
one could have equally well normalized to any of the other two gaugino 
masses. The choice of $M_1$ is only motivated by the expectation that 
it will be the gaugino parameter measured with the smallest error.

\section{Sliding scale imprints in the leading-log RGE Invariants }
\subsection{Models I and II with a sliding 
$SU(2)_{R}$~scale\label{SU2RmodelsResults}}
\begin{figure}[t]
\parbox{8cm}{
\includegraphics[width=8cm]{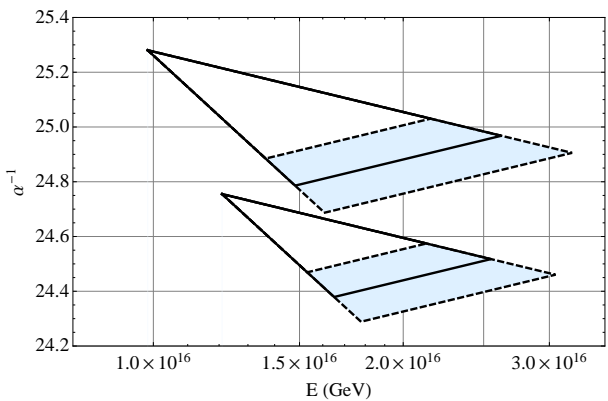}
}
\caption{The MSSM-like non-unification triangle in Models I and III
with $v_{R}=10^{14}$ GeV for two different values of the unknown
soft-SUSY breaking scale ($m_{\rm SUSY}=1$ TeV for the upper one and
$m_{\rm SUSY}=500$ GeV for the lower). The upper sides of the
triangles corresponds to $\alpha_{L}^{-1}$ while the lower-left sides
depict the ``effective'' $\alpha_{Y}^{-1}$ defined as
$\frac{3}{5}\alpha^{-1}_{R}+\tfrac{2}{5}\alpha^{-1}_{B-L}$. The light
blue area surrounding the $\alpha_{S}^{-1}$ line represents the
$1\sigma$ uncertainty in $\alpha_s(M_{Z})$ as given in
\cite{Amsler:2008zzb}. Both triangles move down for lower values of
$v_{R}$, see FIGs.~\ref{fig:ModelIrunning} and
\ref{fig:ModelIIrunning}.}
\label{fig:triangle}
\end{figure}

\begin{figure}[t]
\parbox{8cm}{
\includegraphics[width=8cm]{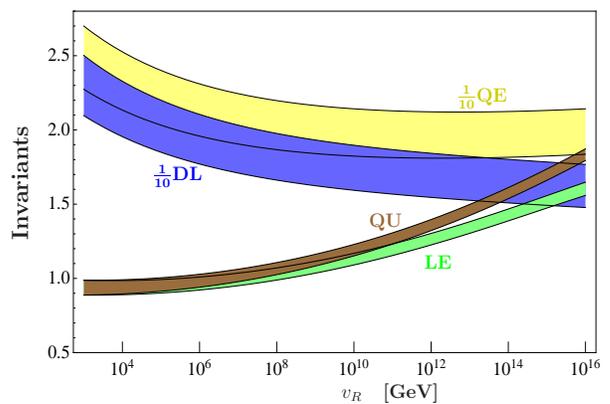}
}
\caption{The $v_{R}$-dependence of the leading-log invariants in
Model~I, c.f., Sect.~\ref{ModelI}. The bands represent the error due
to the non-exact gauge-coupling unification depicted in
FIG.\ref{fig:triangle}. For practical reasons, the numerical values of
the invariants QE and DL have been scaled down by a factor of ten.}
\label{fig:invmodI}
\end{figure}

\begin{figure}[t]
\parbox{8cm}{
\includegraphics[width=8cm]{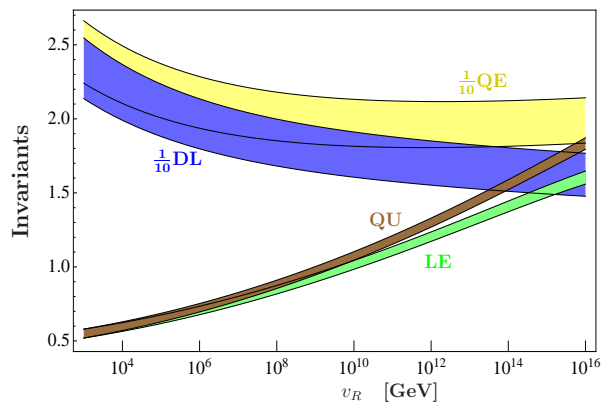}
}
\caption{The same as in FIG.~\ref{fig:invmodI} but for Model~II of
Sect.~\ref{ModelII}. The QU and LE behaviour differs from that in
FIG.~\ref{fig:invmodI} mainly in the low-$v_{R}$ regime.}
\label{fig:invmodII}
\end{figure}

\paragraph*{The method:}
As we have already mentioned in Sect.~\ref{sect:models}, in Models I
and II the sliding nature of the $SU(2)_{R}$ scale makes it impossible
to get an exact unification, in full analogy with the MSSM. Since,
however, this is just about a $2\%$ effect, we shall not attempt to
improve on this by either looking for a suitable set of threshold
corrections or by going beyond the one-loop
approximation\footnote{Indeed, this would be inconsistent as we are
concerned only with the leading-log approximation for the
softs.}. Rather than that, we shall just parametrize our ignorance of
the ``true values'' of the unification scale position and the unified
gauge coupling in terms of a pair of small ``offset'' parameters
scanning over the area of the relevant ``non-unification triangle''
shown in FIG.(\ref{fig:triangle}).  In what follows, we shall to use
the error on $\alpha_S(M_Z)$ given in \cite{Amsler:2008zzb}, $\Delta(
\alpha_S(M_Z)) = 0.002$, which does not take into account the latest
QCD lattice calculations results.
\paragraph*{The results:}
In FIGs.~\ref{fig:invmodI} and~\ref{fig:invmodII} we display the
$v_{R}$-dependence of the RGE invariants in Models I and II due to the
running effects subsumed by Eq.~(\ref{eq:scalar}). The bands
correspond to the error in the gauge-coupling unification inherent to
these settings which, at the leading-log level, can be taken into
account by scanning over the area of the relevant non-unification
triangle, c.f., FIG.~\ref{fig:triangle}. The upper (yellow) band
refers to the combination QE, the (blue) band which at low $v_R$
partially overlaps with QE represents DL, whereas the third (brown)
band is QU and, finally, the lowest (green) band refers to the LE
combination. Note that, for practical reasons, the invariants QE and
DL have been scaled down by a factor of ten. The same colour-code is
adopted in the other figures in this section.

Several comments are in order here: In general, the invariants exhibit
a logarithmic dependence on $v_R$.  For $v_{R}$ close to the MSSM
scale (on the left), the QU and LE invariants overlap. This is
attributed to the enhanced gauge symmetry throughout the whole $m_{\rm
SUSY}$-$M_{\rm G}$ range which makes $m^{2}_{\tilde Q}$ and
$m^{2}_{\tilde U}$ as well as $m^{2}_{\tilde L}$ and $m^{2}_{\tilde
E}$ behave the same, see the LR-stage $c_{i}^{\tilde f}$-coefficients
in TABLE~\ref{tab:coeff}. In the $v_R \to M_{\rm G}$ limit, the mSugra
values of the invariants (modulo the MSSM non-unification) are
reproduced.  Concerning QE and DL, the first thing to notice is that
these invariants tend to increase with $v_{R}$ departing from $M_{\rm
G}$, thus leading to a pattern characteristic to this class of
models. Moreover, they are more sensitive to the initial condition
because the colour-effects in their evolution do not cancel, thus
leading to larger bands.

Naturally, the main difference between FIG.~\ref{fig:invmodII} and
FIG.~\ref{fig:invmodI} is expected in the low-$v_R$ regime where the
effects due to the slight difference in the Model-I and Model-II
spectra are most pronounced and the QU and LE invariants run faster
due to a larger ratio of the coupling constants in the relevant
Eq.~(\ref{eq:fi}).

\subsection{Model~III with sliding $SU(2)_{R}$ and PS scales}
\begin{figure}[t]
\parbox{8cm}{
\includegraphics[width=8cm]{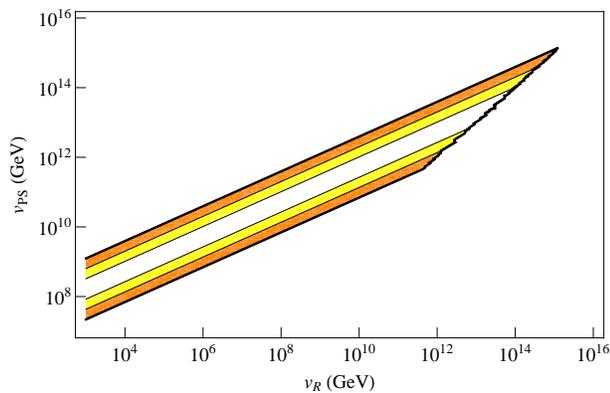}
}
\caption{The correlation of the intermediate symmetry-breaking scales
in Model~III (allowed region coloured). The contours correspond to the
quality of the fit of $\alpha_{S}(M_{Z})$ for each choice of the
Pati-Salam breaking scale $v_{PS}$ an the LR breaking scale $v_{R}$,
within 1$\sigma$ (white area within the coloured band), 2$\sigma$
(yellow) and 3$\sigma$ (orange) of the range quoted in
\cite{Amsler:2008zzb}.}
\label{fig:parametricspacePS}
\end{figure}

\paragraph*{The method:}
In Model~III, the LR and PS intermediate scales can be always adjusted
so that one gets an exact one-loop unification for $v_{R}$ stretching
up to about $10^{14}$ GeV, c.f., FIG.~\ref{fig:parametricspacePS}. 
This is technically achieved by relating the value of the PS scale to 
the value of the LR scale as
\begin{eqnarray}\label{eq:PSLR}
t_{PS}& =& \frac{1}{2}t_{LR} - \frac{1}{12} \Big( 14 t_{SUSY} + 20 t_Z 
\\ \nonumber 
&+& \pi (18 \alpha_S(t_Z)^{-1}-33 \alpha_L(t_Z)^{-1}+15 \alpha_Y(t_Z)^{-1}) 
\Big)
\end{eqnarray}
Here, the $t_{x}$ stand for $ln(m_X)$ as usual. Thus, the main
uncertainty at this level comes from the experimental error in
$\alpha_{S}(M_{Z})$. In what follows, we shall vary $v_{R}$ and
$v_{PS}$ along the constant $\alpha_{S}(M_{Z})$-error trajectories,
namely, within $\pm 1\sigma$, corresponding to the boundaries between
the yellow and white areas within the parameter area depicted in
FIG.~\ref{fig:parametricspacePS}.

\begin{figure}[t]
\parbox{8cm}{
\includegraphics[width=8cm]{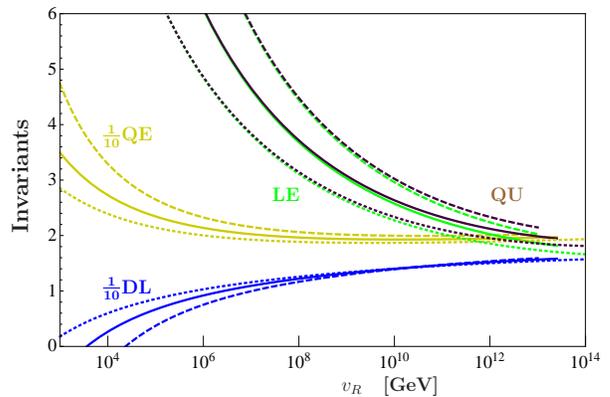}
}
\caption{Intermediate-scale dependence of the RGE invariants
Model~III, see Sect.~\ref{ModelIII}. For each of the four invariants,
the solid curve corresponds to $\alpha_{S}(M_{Z})$ fixed at its
central value and the dashed and dotted lines refer to the $-1\sigma$
and $+1\sigma$ trajectories, respectively;
c.f. FIG~\ref{fig:parametricspacePS}.
\label{fig:invPS}}
\end{figure}

\paragraph*{The results:}
In this case, the intermediate-scale dependence of the leading-log RGE
invariants is yet more pronounced than in Models I and II, c.f.,
FIG~\ref{fig:invPS}. As before, the numerical values of the invariants
QE and DL have been conveniently scaled down by a factor of ten. For
each of the four invariants, the colid curve FIG~\ref{fig:invPS}
corresponds to $\alpha_{S}(M_{Z})$ fixed at its central value and the
dashed and dotted lines refer to the $-1\sigma$ and $+1\sigma$
trajectories, respectively. On the high-$v_{R}$ tail, the different
curves stop at different energies due to the need to respect the
natural $v_{R}<v_{PS}$ hierarchy reflected by the ``diagonal'' cut to
the parametric space in FIG.~\ref{fig:parametricspacePS}.

For all four invariants under consideration, we observe a stronger
$v_{R}$-dependence than in Models I and II. This is namely due to the
extended Pati-Salam running which contributes with larger
$c_{i}^{\tilde f}$-coefficients than the LR stage, c.f.,
TABLE~\ref{tab:coeff}. Moreover, unlike in FIGs.~\ref{fig:invmodI}
and~\ref{fig:invmodII}, three out of four invariants grow with
lowering $v_{R}$ while the fourth one even becomes negative for
$v_{R}$ close to the MSSM scale, thus, again, leading to a very
characteristic pattern.
\subsection{Model~IV with a sliding $U(1)_{R}\times U(1)_{B-L}$ scale}
\begin{figure}[t]
\parbox{8cm}{
\includegraphics[width=8cm]{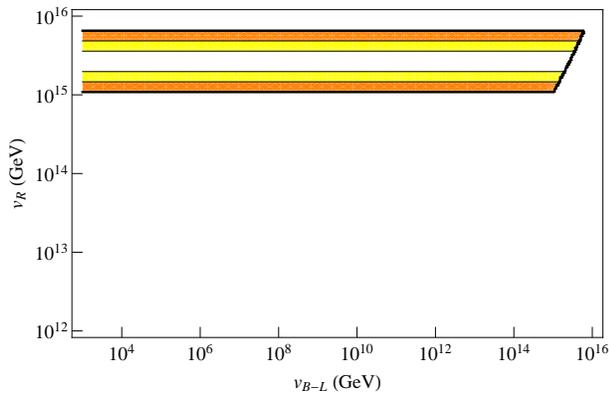}
}
\caption{The parameter space of Model~IV of Sect.~\ref{ModelIV}. The
contours correspond to the different quality of the
$\alpha_{S}(M_{Z})$ fit for different choice of the L-R breaking scale
$v_{R}$, namely, to 1$\sigma$ (white), 2$\sigma$ (yellow) and
3$\sigma$ (orange) values for the range quoted in
\cite{Amsler:2008zzb}.}
\label{fig:parametricspaceMRV}
\end{figure}

\paragraph*{The method:}
Finally, in Model~IV, c.f. Sect.~\ref{ModelIV}, the unification is
exact for any value of the sliding scale $v_{BL}$ below a (constant)
$v_{R}$ in the ballpark of roughly $3\times 10^{15}$ GeV, c.f.,
FIG.~\ref{fig:parametricspaceMRV}.  Thus, as before, the main
uncertainty at this level comes from the experimental error in
$\alpha_{S}(M_{Z})$ which translates into small shifts in $v_{R}$. In
what follows, we shall again vary $v_{BL}$ along the constant
$\alpha_{S}(M_{Z})$-error trajectories, namely, within $\pm 1\sigma$,
corresponding to the boundaries between the yellow and white areas
within the parametric depicted in FIG.~\ref{fig:parametricspaceMRV}.

\begin{figure}[t]
\centering
{\includegraphics[width=8.3cm]{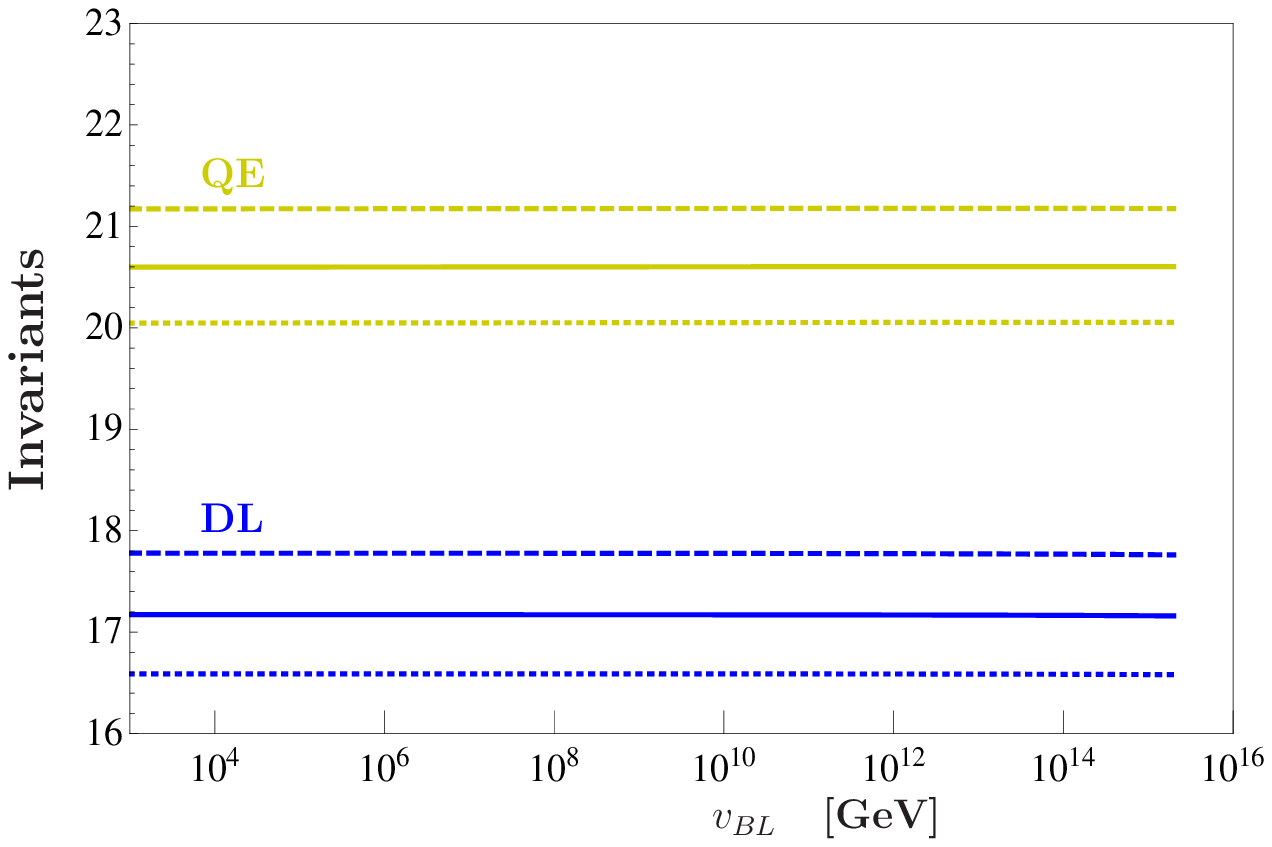}}\\
{\includegraphics[width=8.5cm]{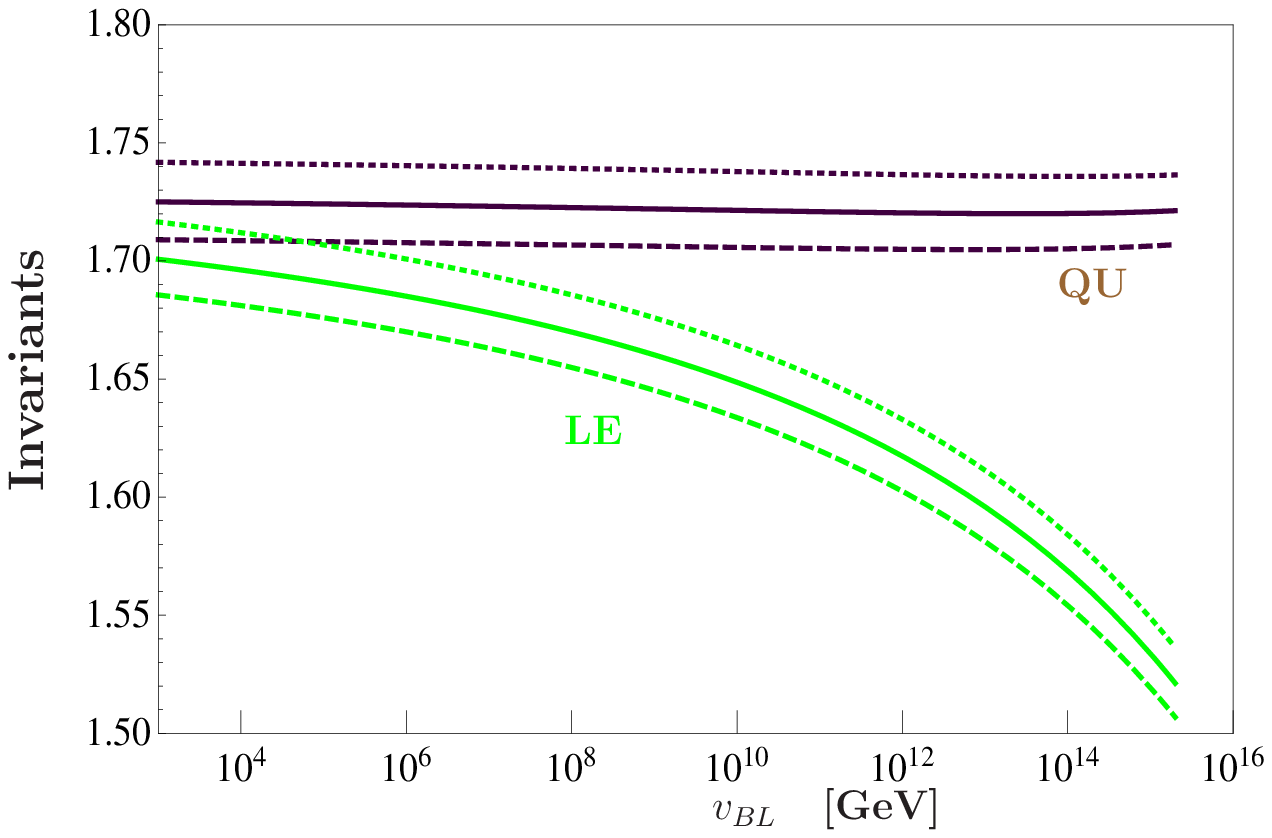}}
\caption{The $v_{BL}$-scale dependence of the RGE invariants in
Model~IV.  For practical purposes the figure has been split into two
panels. For each of the four invariants, the colid curves correspond
to $\alpha_{S}(M_{Z})$ fixed at its central value while the dashed and
dotted lines refer to the $-1\sigma$ and $+1\sigma$ trajectories
corresponding to roughly $v_{R}\sim 2\times 10^{15}$ GeV and
$v_{R}\sim 4\times 10^{15}$ GeV, respectively; c.f.
FIG~\ref{fig:parametricspaceMRV}.}
\label{fig:invMRV}
\end{figure}

\paragraph*{The results:}
In the two panels of FIG.~\ref{fig:invMRV}, the four invariants of our
interest are depicted as functions of $v_{BL}$. As in the case of
Model~III, for each of them the solid line corresponds to the
central-value trajectory in the parametric space of
FIG.~\ref{fig:parametricspaceMRV}, whereas the dashed and dotted
curves refer to the $-1\sigma$ and $+1\sigma$-trajectories,
respectively.

Due to the very special nature of the sliding scale in this setting,
all four invariants exhibit only a very mild $v_{BL}$ dependence, with
the strongest effect of the order of few per cent observed in the LE
case. This is because the $v_{BL}$ scale enters into the soft masses
only through the slight changes in the abelian gauge couplings, which,
however, are overwhelmed by the colour effects in all the other
invariants.  This, however, will make it rather difficult to
distinguish this model from the MSSM, namely because such a
discrimination is efficient only if more than a single invariant
differs significantly from the mSugra value so that the intermediate
scale can be independently constrained from more than a single
quantity.

Finally, let us comment in brief on the case where the $U(1)_{R}\times 
U(1)_{B-L}\to U(1)_{Y}$ breakdown is not triggered by $SU(2)_{R}$ doublets 
like above but by, e.g., $SU(2)_{R}$ triplets. We expect that for such 
models the effects on the invariants would be similar to those expected 
in Model~IV, and certainly smaller than those observed in Models I-III. 

\subsection{Squark and slepton spectra.}
\begin{figure}[h]
\parbox{8.5cm}{
\includegraphics[width=8.5cm]{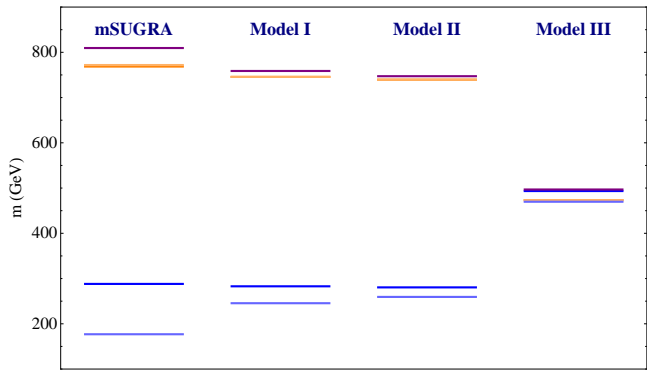}
}
\caption{The MSSM squark and slepton spectra mSugra and Models I, II
and III calculated for the SPS3 benchmark point, i.e. for $m_0 = 90$
GeV and $M_{1/2}=400$ GeV. In all cases, $v_R = 10^{3}$ GeV and
$v_{PS} \sim 10^{7}$ GeV in Model~III . From botton to up the
horizontal lines correspond to: $m_{\tilde{e}^{c}}$ (light blue),
$m_{\tilde{l}}$ (blue), $m_{\tilde{u}^{c}}$ (orange),
$m_{\tilde{d}^{c}}$ (light orange) and $m_{\tilde{q}}$ (purple). We do 
not show the results for model IV in this figure, since they are 
very similar to the mSugra case.}
\label{fig:softmasses}
\end{figure}

In FIG.~\ref{fig:softmasses} we plot the shapes of the MSSM squark
and slepton spectra obtained in mSugra and in Models I, II and III
calculated for the SPS3 benchmark point, i.e. for $m_0 = 90$ GeV and
$M_{1/2}=400$ GeV. This figure is to be understood only as an 
illustrative example of the different spectra generated in our 
different models. For each of the cases, the horizontal lines (bottom
to up) correspond to $m_{\tilde{e}^{c}}$ (light blue), $m_{\tilde{l}}$
(blue), $m_{\tilde{u}^{c}}$ (orange), $m_{\tilde{d}^{c}}$ (light
orange) and $m_{\tilde{q}}$ (purple).  In order to pronounce the
differences, the $v_R$ scale has been in all cases chosen very low,
namely, $v_{R}\sim 10^{3}$ GeV, and consequently $v_{PS}$ in Model~III
is fixed to $v_{PS}\sim 10^{7}$ GeV by gauge unification.  The masses
of the $\tilde{d^{c}}$ and of the $\tilde{u^{c}}$ almost coincide in
all the Models. Models I and II differ from the mSugra case namely by
the smaller splittings observable in the squark as well as in the
slepton masses, which is more pronounced for the latter model.  However, 
the spectrum of Model~III is strongly compressed due to an extended
Pati-Salam stage which makes it rather outstanding.

We decided not to overpopulate the figure by displaying the gaugino
masses which, indeed, are obtained by a simple rescaling
(\ref{eq:gaugino}); the SUSY-to-GUT-scale ratios of the relevant
$\alpha$'s can be inferred from the evolution of the gauge couplings,
c.f. FIGs~\ref{fig:ModelIrunning}, FIGs~\ref{fig:ModelIIrunning},
FIGs~\ref{fig:ModelIIIrunning} and FIGs~\ref{fig:ModelIVrunning}.

\section{Discussion and outlook}
We have studied the leading-log RGE evolution of the MSSM soft SUSY
breaking parameters for four different GUT models with mSugra boundary
conditions. Although all the settings are based on the unified
$SO(10)$ gauge group, they differ at the level of intermediate scale
symmetry groups and/or particle content below the GUT scale. Two of
the models discussed (Models I and II of Sects.~\ref{ModelI} and
\ref{ModelII}), which differ only in their beyond-MSSM field contents,
feature an intermediate left-right symmetry which, at the level of
precision used in this calculation, can be broken to $SU(3)_{c}\times
SU(2)_{L}\times U(1)_{Y}$ of the MSSM almost anywhere between $M_{\rm
G}$ and the soft SUSY-breaking scale. In Model~III (c.f.,
Sect.~\ref{ModelIII}) the sliding nature of the $SU(2)_{R}$-breaking
scale relies on an additional intermediate Pati-Salam
symmetry. Finally, in Model~IV (see Sect.~\ref{ModelIV}), the
left-right symmetry is broken at a relatively high scale, but there is
instead a sliding scale corresponding to the breaking of its
$U(1)_R\times U(1)_{B-L}$ remnant. All models we consider are able to
accommodate the neutrino data by either inverse or linear seesaw.

The extra gauge groups and/or beyond MSSM fields change the evolution
of the soft parameters with respect to the basic mSugra
expectation. The invariant mass combinations we considered are
especially suited to uncover the effects of beyond-mSugra physics on
the SUSY spectra. Remarkably, while invariants contain only a
logarithmic dependence on the new physics scales, their behavior is
{\em qualitatively} different in different models.

In our Models I and II, the invariants LE and QU (c.f.,
Sect.~\ref{sect:invariants}) are always lower than the mSugra limit,
while DL and QE are always larger.~The former is a direct consequence
of the LR symmetry, while the latter reflects mainly the shift in
$\alpha(M_{\rm G})$ the models exhibit with respect to the MSSM
expectation. Moreover, in spite of only a mild difference in the
particle content, the invariants differ quantitatively between Model~I
and Model~II.

In contrast to that, in the Pati-Salem based Model~III, LE and QU are
always larger than in mSugra, with a rather strong dependence on the
$v_R$ scale, namely due to the higher dimensionality of the relevant
multiplets at the Pati-Salam stage. At the same time, in Model~III, DL
is always below the mSugra limit, while QE hardly varies at all as a
function of $v_R$. Finally, Model~IV is an example of how a new scale
can be effectively ``hidden'' from the RGE invariants in special
constructions: Despite containing a new scale potentially as low as
${\cal O}(1)$ TeV, all invariants are always very close to the mSugra
limit in this model. Technically, this is achieved by maintaining the
beta coefficients for the $SU(2)_L$ and $SU(3)_c$ factors as in the
MSSM all the way up to a scale close to $M_{\rm G}$, while the sliding
feature of the $U(1)_R\times U(1)_{B-L}$ breaking scale ``shields''
all invariants from the effects of the new group, with the exception
of LE, which, however, changes only very weakly.

It is especially interesting to compare our results with those
obtained for minimal seesaw models within mSugra. Invariants for
seesaw have been studied previously in
\cite{Buckley:2006nv,Hirsch:2008gh,Esteves:2010ff}. Type-I seesaw adds
only singlets to the MSSM and thus, just like our Model~IV,  can not
be distinguished from the pure mSugra case by means of the invariants
only.  Type-II and type-III seesaw, on the other hand, change the
$b$-coefficients with respect to the MSSM, but do not extend the gauge
group. As a result, for minimal seesaws all four invariants are
larger than their mSugra limit if the seesaw scale is below the GUT
scale, as indicated by neutrino data. Thus, the invariants should
allow to distinguish our $SO(10)$-based Models I to III from type-II
and type-III seesaw.

The RGE invariants are, therefore, good model discriminators, at least
in principle.  However, any attempt to quantitatively determine the
scale of a new physics within a particular scenario must inevitably
address the accuracy of their calculation. Different types of errors
need to be considered here. First, there are the errors from
uncertainties in the values of the input parameters. The largest error
currently stems from the completely unknown $m_{SUSY}$, see
FIG.~\ref{fig:triangle} and Eq.~(\ref{eq:gaugino}).  Once SUSY masses,
indeed, have been measured, this will become irrelevant and the
largest error will, most likely, be $\Delta(\alpha_S)$.

Next, the RGE invariants considered here are calculated to the
leading-log precision only. However, in some cases, important higher
order effects such as genuine 2-loop corrections and 1-loop thresholds
can emerge; for the seesaw, this was studied recently in
\cite{Hirsch:2008gh,Esteves:2010ff}.  Both, 2-loop running and 1-loop
SUSY thresholds can, of course, be taken into account, but the
calculation of the invariants at this level can not be done
analytically. Instead, it requires numerical tools such as, e.g.,
SPheno \cite{Porod:2003um,Porod:2011nf} and SARAH
\cite{Staub:2008uz,Staub:2009bi,Staub:2010jh}.

Probably more important than the above theoretical considerations, 
eventually, will be the fact that the invariants are not directly 
measurable quantities. Conversion of the invariants into the measured 
sparticle masses (or extraction of relevant soft parameters from 
sparticle measurements) requires additional {\em experimental} 
input. In case of the first two generations of sfermions this 
requires at least a reliable measurement of $\tan\beta$ for the 
determination of the D-terms. In addition, at variance with 
the situation in the minimal seeesaw models, the breaking of the extra 
gauge symmetries can potentially produce new D-terms not present in 
the $SU(3)_C\times SU(2)_L\times U(1)_Y$ case. Usually it is assumed 
that any beyond-SM gauge group is broken 
in a ``D-flat'' manner in order to avoid problems with tachyonic sfermions. 
However, since we wish to extract information from sfermion masses 
themselves, it will certainly be prudent to do a combined fit on the 
new parameters instead of simply {\em assuming} D-flatness.

The prospects of measuring sparticle masses at the LHC and, possibly, at the  
future ILC have been studied by many authors, for a detailed 
review see, for example \cite{Weiglein:2004hn}. In general, one 
expects that slepton and gaugino masses, if within the kinematical 
reach of the ILC, can be measured at the per mill level or even better. 
Coloured sparticles, however, might be too heavy to be produced 
at the ILC. At the LHC, the precision with which sparticle masses 
can be measured depends strongly not only on the absolute scale 
of the SUSY masses, but also decisively on the mass ordering of 
the sparticles. If long decay chains such as ${\tilde q}\to \chi^0_2 
q$ with $\chi^0_2 \to {\tilde l}l \to l^{\pm}l^{\mp}\chi^0_1$ are 
available, many SUSY masses can be measured with accuracies 
down to (few) percent. From the detailed studies of  \cite{Weiglein:2004hn}, 
the authors of \cite{Hirsch:2011cw} concluded that the precision 
of ILC+LHC combined would make it possible to see indications for a seesaw 
of either type II or type III for nearly all relevant seesaw scales. 
In an LHC-only analysis, the seesaw scale must be below $10^{14}$ GeV even in favourable circumstances \cite{Hirsch:2011cw} or 
might not leave a trace in the LHC data at all.

Comparing roughly the changes in spectra induced in the seesaw 
models studied in \cite{Hirsch:2011cw} with the changes expected 
in our $SO(10)$ models, we expect that a detailed, numerical 
calculation should be able to probe most, if not all the interesting 
parameter space of our models, if SUSY is found at the LHC and 
precise mass measurements are done with the help of an ILC.

Finally, we would like to mention that the models we have studied 
in this paper have potentially also a rich phenomenology beyond 
the MSSM apart from the invariants. There are the new gauge bosons, 
additional Higgses, additional gauginos/higgsinos, large lepton 
flavour violation and many other effects worth studying.  We plan to 
return to these questions in a future publication.

\section*{Acknowledgments}
The work of V D R is supported by the EU~Network grant UNILHC
PITN-GA-2009-237920.~M. M. is supported by the Marie Curie Intra
European Fellowship within the 7th European Community Framework
Programme FP7-PEOPLE-2009-IEF, contract number PIEF-GA-2009-253119. We
acknowledge support from the Spanish MICINN grants FPA2008-00319/FPA
and MULTIDARK CAD2009-00064 (Con-solider-Ingenio 2010 Programme) and
by the Generalitat Valenciana grant Prometeo/2009/091.

\appendix
\section{One-loop running with U(1) mixing\label{app:mixedrunning}}
In this appendix we give some technical details of the one-loop
evolution of gauge couplings and soft-SUSY-breaking terms in Model IV
of Sect.~\ref{ModelIV} in which extra kinetic mixing effects,
generally present in theories with multiple $U(1)$ gauge factors,
emerge. This, in the approach advocated in, e.g.,
\cite{Fonseca:2011vn}, amounts to extending the notion of the
individual gauge couplings and gaugino masses associated to different
$U(1)$ gauge factors to matrix forms, which, subsequently, complicates
the relevant generalized evolution equations.
\subsubsection{Gauge couplings}
To deal with the effects of the kinetic mixing in cases with more than
a single abelian gauge factor like in Model IV of Sect.~\ref{ModelIV}
it is convenient to work with a matrix of gauge couplings rather than
with each of them individually, which would require an extra RGE for
the kinetic mixing parameters, c.f., \cite{Fonseca:2011vn}. In the
$U(1)_{R}\times U(1)_{B-L}$ case this amounts to defining \be G=
\left(
\begin{array}{cc}
g_{RR} & g_{RX} \\
g_{XR} & g_{XX}
\end{array}
\right)\,.  \ee where $X$ is a shorthand notation for the canonically
normalized $B-L$. The evolution equation can be then written as
\be\label{Ainvrunning} \frac{\rm d}{{\rm d}t} A^{-1}=-\gamma\,, \ee
where $A^{-1}\equiv 4\pi(GG^{T})^{-1}$ and
$t=\frac{1}{2\pi}\log(\mu/\mu_{0})$.  Here we have defined the
relevant matrix of anomalous dimensions by \be \gamma\equiv \sum_{f}
Q_{f}Q_{f}^{T}\,, \ee where the summation is taken over all the chiral
superfields $f$ in the model and $Q_{f}$ denotes a column vector of
$U(1)_{R}$ and $U(1)_{B-L}$ charges of each $f$.

The matching condition between such high-energy gauge couplings
(corresponding to $U(1)_{R}\otimes U(1)_{B-L}$ in the case of our
interest) and the effective-theory one (i.e., $U(1)_{Y}$ of the MSSM)
at scale $t_{0}$ then reads \be {\alpha_{Y}^{-1}}(t_{0})=p_{Y}^{T}\,
A^{-1}({t_{0}})p_{Y}\,, \ee where
$p_{Y}^{T}=(\sqrt{\tfrac{3}{5}},\sqrt{\tfrac{2}{5}})$ are the
coefficients of the hypercharge $Y$ in the space of the $R$- and
$B-L$-charges, namely,
$Y=\sqrt{\frac{3}{5}}T_{R}^{3}+\sqrt{\frac{2}{5}}X$. Thus, one has
\bea
g_Y^{-2}&=&(g_{RR}g_{XX}-g_{RX}g_{XR})^{-2}\left[\frac{3}{5}\left(g_{XX}^{2}+g_{XR}^{2}\right)\right.\\
&&
\left.+\frac{2}{5}\left(g_{RR}^{2}+g_{RX}^{2}\right)-\frac{2}{5}\sqrt{6}\left(g_{RR}g_{XR}+g_{RX}g_{XX}\right)\right]\,.\nn
\eea
\subsubsection{Soft SUSY-breaking terms}
Neglecting for simplicity the Yukawa couplings and also the
``Trace-terms'' (denoted by $\cal S$ in \cite{Martin:1993zk}), which
in mSugra yield only sub-leading correction to the leading-log
approximation used in this work, one can write the generalized
evolution equation including the effects of the $U(1)$ mixing
\cite{Fonseca:2011vn} as \be\label{scalarevolution} \frac{\rm d}{{\rm
d}t}\tilde m^{2}_{f}= -\frac{1}{\pi} Q^{T}_{f}G M
M^{\dagger}G^{T}Q_{f}\,, \ee where $G$ is the matrix of gauge
couplings, $M$ is the gaugino mass matrix and $t=\frac{1}{2\pi}\log
\mu/\mu_{0}$. This is to be solved together with the gauge coupling
(\ref{Ainvrunning}) and gaugino evolution equations. The latter reads
at one loop \be\label{gauginoevolution} \frac{\rm d}{{\rm d}t}M=
\frac{1}{8\pi}\left(MG^{T}\gamma G+G^{T}\gamma G M\right)\,.  \ee The
simplicity of the system (\ref{Ainvrunning}), (\ref{scalarevolution})
and (\ref{gauginoevolution}) and, in particular, the flavour-diagonal
mSugra initial condition, admits to write the general solution in a
closed and compact form \bea A^{-1}(t)&=&A^{-1}(t_{0})-\gamma
(t-t_{0})\,,\\ (G^{-1T}MG^{-1})(t) &=&\frac{1}{4\pi}
\alpha^{-1}_{G}M_{1/2}\,, \eea and, in particular, \bea && \tilde
m_{f}^{2}(t)-\tilde m_{f}^{2}(t_{0})=2M_{1/2}^{2}\alpha_{G}^{-2}\times
\\ &&\qquad \qquad
Q^{T}_{f}A^{-1}_{0}\left[\gamma^{-1}-A^{-1}A_{0}\gamma^{-1}A_{0}A^{-1}\right]A^{-1}_{0}
Q_{f}\,,\nn \eea where $A_{0}\equiv A(t_{0})$ and $A\equiv A(t)$.

\end{document}